\documentclass[12pt]{article}
\pdfoutput=1
\usepackage{setspace}

\usepackage{graphicx}
\usepackage{epstopdf}
\usepackage[body={17.5cm, 21cm},right=2cm]{geometry}
\usepackage{amssymb}
\usepackage{amsmath}
\usepackage{dcolumn}% Align table columns on decimal point

\usepackage{graphicx}
\usepackage{epstopdf}
\usepackage{amsmath}
\usepackage{amsfonts}
\usepackage{amssymb}
\usepackage[usenames]{color}
\usepackage{mathrsfs}
\usepackage{array}

\definecolor{purple}{rgb}{0.78,0.18,0.77}

\usepackage{multirow}
\usepackage{psfrag}
\pdfoutput=1

\usepackage[colorlinks,bookmarks]{hyperref}
\definecolor{linkblue}{rgb}{0,0,0.8}
\definecolor{linkgreen}{rgb}{0,0.5,0}

\hypersetup{pdfpagemode=UseNone, pdfstartview=FitH, linkcolor=linkblue, %
            citecolor=linkgreen, urlcolor=linkblue}
            
\usepackage[numbers,sort&compress]{natbib}

\newcommand\nn{\nonumber}

\newcommand\eea{\end{eqnarray}}
\newcommand\bea{\begin{eqnarray}}

\def\beq{\begin{equation}}
\def\eeq{\end{equation}}

\def\eps{\varepsilon}

\def\l{\left(}
\def\r{\right)}

\newcommand{\be}{\begin{equation}}
\newcommand{\ee}{\end{equation}}
\newcommand{\ba}{\begin{align}}
\newcommand{\ea}{\end{align}}
\newcommand{\bg}{\begin{gather}}
\newcommand{\eg}{\end{gather}}
\newcommand{\bseq}{\begin{subequations}}
\newcommand{\eseq}{\end{subequations}}

\newcommand{\vk}{\vec{k}}
\newcommand{\vkp}{\vec{q}}

\newcommand{\vq}{\vec{q}}
\newcommand{\vp}{\vec{p}}

\newcommand{\knl}{k_{\rm NL}}
\newcommand{\kmin}{k_{\rm min}}
\newcommand{\vkdq}{\vk\cdot\vq}
\newcommand{\vl}{\vec{\ell}}

\newcommand{\q}{\vec{q}}
\newcommand{\p}{\vec{p}}

\newcommand{\invMpc}{\,h\, $Mpc$^{-1}\,}

\begin{document}

\vspace{5mm}
\vspace{0.5cm}
\begin{center}

\def\thefootnote{\fnsymbol{footnote}}

{\Large \bf The 2-loop matter power spectrum \\[0.5cm]  and the IR-safe integrand}
\\[0.8cm]

{\large John Joseph M. Carrasco$^{1}$, Simon Foreman$^{1,2}$,\\[0.5cm] 
Daniel Green$^{1,2}$, and Leonardo Senatore$^{1,2,3}$}
\\[0.5cm]

{\normalsize { \sl $^{1}$ Stanford Institute for Theoretical Physics and Department of Physics, \\Stanford University, Stanford, CA 94306}}\\
\vspace{.3cm}

{\normalsize { \sl $^{2}$ Kavli Institute for Particle Astrophysics and Cosmology, \\ Stanford University and SLAC, Menlo Park, CA 94025}}\\
\vspace{.3cm}

{\normalsize { \sl $^{3}$ CERN, Theory Division, 1211 Geneva 23, Switzerland}}\\
\vspace{.3cm}

\end{center}

\vspace{.8cm}

\hrule \vspace{0.3cm}
{\small  \noindent \textbf{Abstract} \\[0.3cm]
Large scale structure surveys are likely the next leading probe of cosmological information. It is therefore crucial to reliably predict their observables. The Effective Field Theory of Large Scale Structures (EFTofLSS) provides a manifestly convergent perturbation theory for the weakly non-linear regime, where dark matter correlation functions are computed in an expansion of the wavenumber $k$ over the wavenumber associated to the non-linear scale $\knl$. To push the predictions to higher wavenumbers, it is necessary to compute the 2-loop matter power spectrum. 
For equal-time correlators, exactly as with standard perturturbation theory, there are IR divergences present in each diagram that cancel completely in the final result.   
We develop a method by which all 2-loop diagrams are computed as one integral, with an integrand that is manifestly free of any IR divergences.
This allows us to compute the 2-loop power spectra in a reliable way that is much less numerically challenging than standard techniques. 
We apply our method to scaling universes where the linear power spectrum is a single power law of $k$, and where IR divergences can particularly easily  interfere with accurate evaluation of loop corrections if not handled carefully.
We show that our results are independent of  IR cutoff and, after renormalization, of the UV cutoff, and comment how the method presented here naturally generalizes to higher loops.
\noindent 
}
 \vspace{0.3cm}
\hrule
\def\thefootnote{\arabic{footnote}}
\setcounter{footnote}{0}

\vspace{.8cm}

\iffalse
\section{Outline}
\begin{enumerate}
\item Introduction
\item  expressions $P_{51}$ tex.

\item IR divergences
\item Example of 1-loop 
\item  our tricks and Final form 2-loop
\item Application to scaling universe $n=-3/2$
\begin{enumerate}
\item estimates on scalings 
\item UV divergences and counterterms
\item results for the divergent term and for the finite terms (with $k^{3/2}$ having no counterterm)
\item comparison with other codes
\end{enumerate}
\item JJ should please to provide: 
\begin{enumerate}
\item High-precision data for scaling universe n=-3/2 with quasi montecarlo
\item High-precision data for scaling universe n=-3/2 with Vegas montecarlo
\item High-precision data for scaling universe n=-5/2 with quasi montecarlo
\item High-precision data for scaling universe n=-5/2 with Vegas montecarlo
\item Same as above for RegPT code
\item Same as above for Whyte's code (which I guess it fails immediately, so we say it fails).
\end{enumerate}
\end{enumerate}

\fi

\clearpage
\section{Introduction}

Large scale structures have the potential of becoming the next leading cosmological probe for the initial conditions of the universe.  Our knowledge of the initial conditions scales as the cube of the maximum wavenumber in observations that we can theoretically predict.  This is a tremendous amount of potential data.

On short scales density perturbations have become non-linear and have undergone collapse. A description of this regime likely necessitates the use of $N$-body numerical simulations, with analytical techniques providing guidance.  The situation is very different on large scales, where  non-linearities are weak  and  an analytical treatment should be possible.  The recently formulated Effective Field Theory of Large Scale Structures (EFTofLSS)~\cite{Baumann:2010tm,Carrasco:2012cv} is the theory that allows us to consistently make predictions\footnote{Effective field theories, by definition, offer consistent perturbative predictions for their scales of relevance, whose accuracy is determined by the perturbative order of calculation up to nonperturbative ambiguity.  This is simply the old-recognition that the identification of important operators can be systematized through consideration of symmetry and characteristic scales.  } for correlation functions at a certain wavenumber $k$ in an expansion in $\delta\rho(k)/\rho$, or equivalently $k/\knl$, with $\knl$ being the scale at which nonlinear corrections can no longer be treated perturbatively. A standard heuristic is that $\knl$ should correspond to the scale at which the matter overdensity, $\delta\rho(k)/\rho$, becomes of order unity. However, order-one factors arising from the details of the theory will alter this estimate, so that the precise scale at which perturbation theory will fail cannot be precisely known until the computations are performed.  Indeed we find that the EFT can reach much higher wavenumbers than previous estimates for the breakdown of perturbative description.   Since we are interested in computing correlation functions for $\delta(k)\ll 1$ or equivalently $k\ll \knl$, this perturbation theory is manifestly convergent~\footnote{ By this we simply mean that each higher loop order has a better scaling in $k/\knl$ than previous loop orders, so that, until nonperturbative corrections become important, each order in perturbation theory is smaller than the previous one for $k\lesssim \knl$. Notice that this is true only when one scale is involved in the problem. In the current universe there are two additional scales, the matter-radiation equality scale and the scale of Baryon Acoustic Oscillations~(BAO), and so there is the potential of additional large factors ruining the perturbative expansion.  This is indeed the case in the current universe when one deals with IR-unsafe quantities or when one is concerned with reconstructing the BAO. If one is interested in these quantities, a resummation of the IR effect need to be performed in order for the perturbative expansion to be just in powers of $k/\knl$.}.   Indeed, by definition, $\knl$ is the wavenumber for which the EFT stops converging. 

 The effective theory differs from standard, normally used, perturbation theories~\cite{Bernardeau:2001qr} by additional terms in the fluid equations of motion, such as the speed of sound, viscosity, and stochastic pressure.  In addition, loops are performed formally with a cutoff and these additional terms of the EFT have the role of canceling the cutoff dependence for physical observables and to add a finite contribution. After this step, the theory is said to be renormalized, and it is only at this point that the expansion in $k/\knl$ is manifest~\cite{Carrasco:2012cv}. The EFTofLSS was developed and applied to simulation data in~\cite{Carrasco:2012cv}, where it was explicitly shown how the additional terms  of the EFT remove the cutoff dependence of the loops and how the theory is renormalized. Since the additional terms such as the speed of sound are incalculable within the effective theory, they need to be either measured from observed data, or extracted from $N$-body simulations. Both approaches were developed in~\cite{Carrasco:2012cv}, where it was shown that the 1-loop prediction of the EFTofLSS is in agreement~\footnote{The order of magnitude of the UV effects, which are exactly included in the EFTofLSS formalism, was estimated in~\cite{ScocUV}. The size that ends up appearing in explicit comparisons of EFTofLSS calculations to data is larger than the estimates by a factor of a few. However there is no particular tension between the two results given the fact that the estimates in~\cite{ScocUV} neglect ${\cal O}(1)$ factors. Indeed, the precise UV effects were accurately measured from short-scale correlation functions in~\cite{Carrasco:2012cv}, and found to agree with what is needed to match the power spectrum up to $k\sim 0.24\invMpc$ with great accuracy.  } 
 with the dark matter power spectrum to within a percent up to the relatively high scale of $k\sim 0.24 \invMpc$. The remarkable agreement with observations, the improvement with respect to other currently available techniques (see for example~\cite{Carlson:2009it}), and the self-consistency of the different methods of extracting the parameters of the EFT, give very strong evidence that the EFT is the correct language to make theoretical predictions for LSS. This becomes even more evident when one tries to make predictions for other toy, `scaling', universes where the initial power spectrum follows a simple power law. In this set up, all techniques other than the EFTofLSS are unable to make predictions, while the EFT can~\cite{senatoretalk,Pajer:2013jj}. By using the renormalization techniques developed in~\cite{Carrasco:2012cv}, Ref.~\cite{Pajer:2013jj} has explicitly verified that indeed the EFTofLSS is able to make predictions for the scaling universes.

From the perspective of the EFTofLSS, the former techniques are missing important terms in the equations of motion.  Once these terms are included, perturbation theory may converge for much larger $k$ than previously thought.  Given the importance of increasing the window of modes over which we can reliably predict observables, and given the encouraging results of~\cite{Carrasco:2012cv}, it is tempting to perform calculations beyond one-loop. While conceptually straightforward, the next-to-leading order contribution to the power spectrum, from 2-loop diagrams, happens to be technically subtle. Apart for the presence of counterterms that need to be evaluated at lower loop order, the calculation of the 2-loop contribution is actually exactly the same as in Standard Perturbation Theory (SPT)~\cite{Bernardeau:2001qr}. The calculation is numerically challenging. There are four diagrams, denoted as $P_{51},\;P_{42}$, $P_{33}^{\rm (I)}$ and $P_{33}^{\rm (II)}$,  each one involving a  5-dimensional integral. One of the main difficulties is that  each diagram has infrared (IR) divergences as some of internal momenta go to zero. These must cancel between diagrams given infinite precision, but finite resources make it easy to generate spurious numerical errors.  It is a peculiar fact associated to the nice IR behavior of the standard real-universe linear power spectra that such spurious issues, involving the computational delicacy of higher-loop cancellation, have largely gone unnoticed until now -- although as we will discuss such numerical noise can very much be a real issue with noticeable consequences if not correctly addressed.  These issues become more evident in the case of scaling universes that we treat here.

  As proven in~\cite{Jain:1995kx,Scoccimarro:1995if}, in equal-time matter correlation functions, all IR divergences must cancel when we sum together all contributing diagrams, provided that the linear matter power spectrum~$P_{11}$ grows in the IR more slowly than $k^{n}$ with $n>-3$  (see \cite{Peloso:2013zw} for a recent discussion). The physical reason behind this is the following. If we consider two short modes, they will see long modes as a background, and they will be carried over by the velocity of the  long modes, the so-called bulk flows. By the equivalence principle, the dynamics of the short modes is sensitive to long modes only through second derivatives of the metric, the so-called tidal forces, which go as $k^3 P_{11}(k)$. If $n<-3$, these tidal forces grow larger for longer wavelengths, leading to IR divergent correlation functions. For equal-time correlation functions this is all that there is to it, and so IR divergences are guaranteed to be absent for $n>-3$ by this argument. For non-equal time correlation functions, there is an additional effect through which IR modes appear. This is in inducing a relative displacement between the fluctuation evaluated at the earlier time and the one at the later time. Since the power spectrum of velocities goes as $k^3 P_{11}(k)/k^2$, these IR divergences at non-equal times will cancel only for $n>-1$. 

In this paper we will focus on equal time correlations with $n>-3$, where IR divergences are guaranteed to cancel. However, the fact that this cancellation happens only when summing over all the diagrams introduces difficulties at the numerical level. In fact, in the currently available codes that perform the two-loop calculation for SPT and that we have tested~\cite{Carlson:2009it,Taruya:2012ut}, each diagram is evaluated separately. This means that each diagram is IR divergent and so needs to be regularized in the IR. Ideally, if the integrals are computed exactly, the large IR dependent part will cancel and we will be left with a smaller, IR-independent part. However, this means that the integrals need to be evaluated to a much greater precision than what is really necessary, because the leading part will actually cancel. To be explicit, if we wish to compute the two-loop power spectrum with~$10^{-2}$ relative precision, and the IR dependent part of each diagram is $10^2$ larger than the finite part, we need to evaluate each diagram to $10^{-4}$ precision. This is clearly a numerical challenge and a waste of resources. Even worse, as we will explain, in an expansion of the IR-independent part in powers of $k/\knl$, the leading IR-independent part is actually degenerate with the effect of a counterterm from the EFTofLSS, the speed of sound term. This means that we are not interested in computing the leading IR-independent part, but actually the next subleading part. This raises the need for higher, numerical precision, making the challenge even harder.

We find that the currently available 2-loop codes that we have tested~\cite{Carlson:2009it,Taruya:2012ut} are  not precise enough for scaling universes, probably for the aforementioned reasons~\footnote{We stress
the codes we will use here for comparison are not designed for scaling universes. It might well be that some of these IR-issues are numerically irrelevant for these codes in the case of the true universe.}. In contrast we present and implement what we call the IR-safe global integrand. The main idea will be the following: since IR divergences cancel between diagrams, we will bring the integrands of all the four diagrams under one integral. Additionally, in order to cancel IR divergences that happen in different regions of the phase space (for example as the internal momentum goes to zero or to minus the external momentum), we will perform some changes of variables in the integration coordinates so that all the IR divergences happen only as the internal momenta go to zero. In this way, {\it the resulting integrand will be IR convergent}, the cancellation of IR divergences will not rely on numerical precision, but it will happen directly at the level of the integrand and therefore will be automatic. Consequently, the integration computes the {\it IR-finite} part directly, so that the precision with which the integrals are evaluated is the precision of the IR-independent part.  

Because of the numerical challenge, we will apply and test our approach on a scaling universe. This is interesting and useful also because simple dimensional analysis allows us to estimate the parametric result of the integral, allowing a direct verification of success. Furthermore, scaling universes do not have a property of the current universe that gives the illusion that a code, when applied on the true universe, is IR-safe. In fact, in the current universe, for scales longer than the equality scale $k_{eq}$, the power spectrum scales as $k^1$, which makes each loop diagram much less  IR divergent. Because of this, it is possible that some numerical approaches only approximately cancel the IR divergences, but do so well enough for scales longer than the equality scale, so that most of the contribution for a given high-$k$ mode comes from $k_{eq}$. This result would be naively IR-independent, but it is still not the correct answer, as, for correlations evaluated at equal times, every $k$ mode receives contribution only from its neighboring or higher-$k$ modes.  In a separate paper~\cite{nextpaper}, we will apply the IR-safe global integrand to give the 2-loop prediction of the power spectrum from the EFTofLSS, where we will see that the EFTofLSS predicts the power spectrum with percent precision up to roughly $k\sim 0.5\,h\,$Mpc$^{-1}$.

\section{The IR-safe global integrand}
In this section we will be interested only in the loop terms, so we will therefore momentarily neglect the EFT counterterms. This means that the expressions we write in this section are the same\footnote{ Strictly speaking, for the EFTofLSS, one should cut off the UV of the linear power spectrum, e.g.\ using a Gaussian as done in~\cite{Carrasco:2012cv} or with a sharp cutoff $P_{11,EFT} (k)= \Theta({\Lambda-k}) P_{11}(k)$, and renormalize by taking $\Lambda\to\infty$.  As it would muddy the applicability of the integrand discussion for other uses of SPT integrals, and only has mild effects for the measurements taken of the 2-loop power spectra calculated here, we neglect this in the current paper where we focus instead on cancelling bad IR behavior at the integrand level.  One could also consider introducing an explicit IR cutoff on $P_{11}(k)$, e.g.  $\Theta(k-\kmin)$, but as the scaling linear spectra are well defined for $k>0$ we do not do so either.  Rather, we simply establish the bounds of integration from $\kmin$ to $\Lambda$ for all radial wavenumber loop-level integrals -- which should make it transparent how to incorporate the IR-safe integrand we present for any numerical evaluation that is not necessarily EFT focussed, e.g.~\cite{Carlson:2009it,Taruya:2012ut}.} as in SPT. Furthermore, when necessary to use explicit linear power spectra, we will consider scaling universes where the linear power spectrum, defined as $P=P_{11}$, is taken  to be
\be
P(k)=\frac{1}{\knl^3}\left(\frac{k}{\knl}\right)^n\ ,
\ee
with the tilt, $n$, being some number.

\subsection{1-loop Integrand}

The 1-loop matter power spectrum is often written as
\be
P_{\rm 1-loop} = P_{13}+P_{22}
\ee
where
\begin{align}
\nn
P_{13}(k) &= \int \frac{d^3q}{(2\pi)^3} \; p_{13}(\vk,\vq)
= \int \frac{d^3q}{(2\pi)^3} \; 6 P(k) P(q) \; F_3^{(\rm s)}(\vk,\vq,-\vq)\ , \\
P_{22}(k) &= \int \frac{d^3q}{(2\pi)^3} \; p_{22}(\vk,\vq)
= \int \frac{d^3q}{(2\pi)^3}\; 2 P(q) P(|\vk-\vq|) \left[ F_2^{(\rm s)}(\vq,\vk-\vq) \right]^2 \ ,
\end{align}
with the kernels $F_2^{\rm (s)}(\vq,\vk-\vq)$ and $F_3^{\rm (s)}(\vk,\vq,-\vq)$ given by
\begin{align}
\nn
F_2^{\rm (s)}(\vq,\vk-\vq) &=
\frac{k^2 \l 7\vkdq + 3q^2 \r - 10 \l \vkdq \r^2}{14q^2|\vk-\vq|^2}, \\
\nn
F_3^{\rm (s)}(\vk,\vq,-\vq) &=
\frac{1}{|\vk-\vq|^2} \left[ \frac{5k^2}{126} - \frac{11\vkdq}{108}
+ \frac{7( \vkdq )^2}{108k^2} - \frac{k^2 ( \vkdq )^2}{54q^4}
+ \frac{4( \vkdq )^3}{189q^4} \right. \\
\nn
&\qquad \left. - \frac{23k^2 \vkdq}{756q^2}
+\frac{25( \vkdq )^2}{252q^2} - \frac{2 ( \vkdq )^3}{27k^2q^2} \right] \\
\nn
&\quad +\frac{1}{|\vk+\vq|^2} \left[ \frac{5k^2}{126} + \frac{11\vkdq}{108}
- \frac{7(\vkdq)^2}{108k^2} - \frac{4k^2(\vkdq)^2}{27q^4}
- \frac{53(\vkdq)^3}{189q^4} \right. \\
&\qquad \left. + \frac{23k^2 \vkdq}{756q^2} 
- \frac{121(\vkdq)^2}{756q^2} - \frac{5(\vkdq)^3}{27k^2q^2}\right].
\end{align}
As $q\to 0$, the integrands have~\cite{Goroff:1986ep} the following asymptotic behavior (ignoring a common factor of $(2\pi)^{-2}$ and including a $q^2$ factor from the integration in spherical coordinates):
\begin{align}
\nn
q^2 \, p_{13}(\vk,\vq) &\underset{q\to 0}{\sim} 
 - k^{n+2}\mu^2 q^n + \mathcal{O}(q^{n+2}) \ , \\
q^2 \, p_{22}(\vk,\vq) &\underset{q\to 0}{\sim} 
  \frac{1}{2}  k^{n+2} \mu^2 q^n
+ \mathcal{O}(q^{n+1}) \ ,
\label{eq:1looplead}
\end{align}
where $\mu=(\vkdq)/(kq)$. We will discuss the $\mathcal{O}(q^{n+1})$ term momentarily. One sees that the leading IR divergence in $p_{22}$ does not cancel the one in $p_{13}$. In fact, there is another IR divergence in $p_{22}$ as $\vec q\to \vec k$, associated with sending the internal momentum $\vec q-\vec k$ to zero:
\be
q^2 \, p_{22}(\vk,\vq) \underset{\vq\to \vk}{\sim}  \frac{1}{2} k^n
\frac{(\vk\cdot[\vk-\vq])^2}{|\vk-\vq|^2} |\vk-\vq|^n
+ \mathcal{O}(|\vk-\vq|^{n+1})  \ .
\ee
Remarkably, if we sum the leading IR-divergent contributions, they cancel and we are left with an IR independent result. This is guaranteed to happen from the Galilean invariance of the equations of motion~\cite{Jain:1995kx,Scoccimarro:1995if}, and is also guaranteed for all subleading IR divergences. However, the presence of the IR divergences in each of the diagrams makes the calculation numerically much more  challenging. If one is to evaluate $P_{13}$ and $P_{22}$ separately and then sum them up at the end, one needs to put an IR cutoff, compute two very large numbers, and then add the partial results only to obtain a subleading piece. This affects the precision of the result: a certain relative precision in $P_{13}$ or in $P_{22}$ translates into a much worse relative precision in the final result for~$P_{\text{1-loop}}$.

One might think that this problem appears only in the toy scaling universe that we are considering. In the true universe, $P_{11}\sim k$ for $k\lesssim k_{eq}$, making the IR divergence of $P_{13}$ and $P_{22}$ vanish. Still, this means that the result of $P_{13}$ and $P_{22}$ is dominated by the contribution of modes of order $k_{eq}$, whose contribution will cancel for modes $k\gg k_{eq}$. While at 1-loop the numerical challenge is relative, and one can afford such an inefficiency, the challenge becomes much harder for the 2-loop case that we consider next.
  
Since IR divergences are guaranteed to cancel, it would be ideal to compute the 1-loop (or any $n$-loops) power spectrum without IR divergences ever appearing. Ideally, one would like to have an {\it integrand} that does not have any IR divergence. Clearly, given that the IR cancellation happens between $P_{22}$ and $P_{13}$, the only possibility is to write the two as an integration of a {\it single} integrand:
 \be
 P_{\text{1-loop IR-safe guess}}=\int \frac{d^3q}{(2\pi)^3} \left[p_{13}(\vec k,\vec q)+p_{22}(\vec k,\vec q) \right].
 \ee
 This is not enough though, as $p_{22}$ diverges in a symmetric way, both as $\vq\to 0$ and as $\vec q\to \vec k$. However, it is possible to perform the following change of variables on $P_{22}$ in order for it not to have any divergence at $\vec q\to \vec k$.  The idea is split the region of integration in order to isolate each singular term, and to map the singularity at $\vq\to \vec k$ to one at $\vq\to 0$. We write~\footnote{One can derive this more simply by un-doing the Dirac $\delta$-function of momentum conservation:
 \bea
P_{22} &=& \int \frac{d^3q}{(2\pi)^3} \; p_{22}(\vk,\vq)
=\int \frac{d^3q}{(2\pi)^3} \int \frac{d^3p}{(2\pi)^3}
\; \delta_{\rm D}(\vq+\vp+\vk)\;  2 P(q) P(p) \left[ F_2^{(\rm s)}(\vq,\vp) \right]^2\\
\nn
&=& 2\int \frac{d^3q}{(2\pi)^3} \int \frac{d^3p}{(2\pi)^3}
\; \delta_{\rm D}(\vq+\vp+\vk)\;  \Theta(p-q) \, 2 P(q) P(p) \left[ F_2^{(\rm s)}(\vq,\vp) \right]^2\\
\nn
&=&2 \int \frac{d^3q}{(2\pi)^3}  \, \Theta(|\vk-\vq|-q) 2 P(q) P(|\vk-\vq|) \left[ F_2^{(\rm s)}(\vq,\vk-\vq) \right]^2 
= 2 \int \frac{d^3q}{(2\pi)^3}\; p_{22}(\vk,\vq) \Theta(|\vk-\vq|-q) \, ,
\eea
where in the first passage of the second line we have used the symmetry of the integrand under exchange of $\vq$ and~$\vp$.}
\begin{align}
P_{22} &= \int_{|\vq|<|\vk-\vq|} \frac{d^3q}{(2\pi)^3}\; p_{22}(\vk,\vq)
+ \int_{|\vq|>|\vk-\vq|} \frac{d^3q}{(2\pi)^3}\; p_{22}(\vk,\vq) \\
\nn
&= \int_{|\vq|<|\vk-\vq|} \frac{d^3q}{(2\pi)^3}\; p_{22}(\vk,\vq)
+ \int_{|\vec{\tilde{q}}|<|\vk-\vec{\tilde{q}}|} \frac{d^3\tilde{q}}{(2\pi)^3}\; p_{22}(\vk,\vk-\vec{\tilde{q}}) \\
\nn
&= 2 \int_{|\vq|<|\vk-\vq|} \frac{d^3q}{(2\pi)^3}\; p_{22}(\vk,\vq) = 2 \int \frac{d^3q}{(2\pi)^3}\; p_{22}(\vk,\vq) \Theta(|\vk-\vq|-\vq) \,
\end{align}
where in the first passage we have split the region of integration, in the second we have changed integration coordinates from $\vq$ to ${\vec{\tilde{q}}}=\vk-\vq$, and in the third we have relabelled $\vec {\tilde{q}}$ as $\vq$ and used the property $p_{22}(\vk,\vq)=p_{22}(\vk,\vk-\vq)$, obtaining a factor of 2.

In this form, $P_{22}$ has an IR divergence only as $\vq\to 0$, and we can see from Eq.~(\ref{eq:1looplead}) that it will cancel with the corresponding term in $p_{13}$. The final important point is that $p_{22}$ also has a {\it subleading} divergence (the $\mathcal{O}(q^{n+1})$ piece in Eq.~(\ref{eq:1looplead})), with the following behavior:
\beq
q^2 \, p_{22}(\vk,\vq) \underset{q\to 0}{\sim} 
  \frac{1}{2}  k^{n+2} \mu^2 q^n
+ \frac{1}{14} k^{n+1} \mu \, (6+8\mu^2-7n\mu^2) q^{n+1} + \mathcal{O}(q^{n+2}) \ .
\eeq
This term, being odd in $\mu$, disappears once the angular integral is performed, and can also be eliminated in the integrand by symmetrizing in $\vq$ and $-\vq$. Thus, we can write the final IR-safe $P_{\text{1-loop}}$ as
\bea
\nn
P_{\text{1-loop IR-safe}} &=& \int \frac{d^3q}{(2\pi)^3} \left[ p_{13}(\vk,\vq) 
+ p_{22}(\vk,\vq) \Theta(|\vk-\vq|-q)
+ p_{22}(\vk,-\vq) \Theta(|\vk+\vq|-q) \right] \\
\nn
&=& \int \frac{d^3q}{(2\pi)^3}
\left[ 6 P(k) P(q) \, F_3^{(\rm s)}(\vk,\vq,-\vq) \right. \\
\nn
&&\qquad\qquad \left.
+ 2 P(q) P(|\vk-\vq|) \left[ F_2^{(\rm s)}(\vq,\vk-\vq) \right]^2 \Theta(|\vk-\vq|-q) \right. \\
\nn
&&\qquad\qquad \left.
+ 2 P(q) P(|\vk+\vq|) \left[ F_2^{(\rm s)}(-\vq,\vk+\vq) \right]^2 \Theta(|\vk+\vq|-q) \right]\ . \\
\eea

There are two advantages to writing the 1-loop integral in this way. A trivial one is that we have to do one integration instead of two, which is less time-consuming. More importantly, a second one is that  {\it the integrand has no divergences as $q\to 0$} for any $n>-3$. This is guaranteed to be so: the sum of all IR divergences must cancel in the integral, and since we have only one integrand and the internal momenta go to zero only for $\vq\to 0$, the integrand itself must be finite, or at least integrable, for $n>-3$. As we have just explained, this makes an accurate numerical integration much easier to perform. These manipulations of the integrand,  are not so important for the (relatively straightforward) 1-loop integrals, but become essential when one tries to evaluate the much more challenging 2-loop diagrams.

\subsection{2-loop Integrand}
 
Now, let us examine the 2-loop integrals. They are usually written as 
\beq
P_{\text{2-loop}} = P_{15}+P_{24}+ P_{33}^{\rm (I)}+ P_{33}^{\rm (II)}
\eeq
where
\begin{align}
\label{eq:2loopints}
\nn
P_{51}(k) &= \int \!\! \frac{d^3 p}{(2\pi)^3} \int \!\! \frac{d^3 q}{(2\pi)^3} \, 30
F_5^{(\rm s)}(\vk,\vkp,-\vkp,\vp,-\vp) P_{11}(k) P_{11}(q) P_{11}(p)\ , \\
\nn
P_{42}(k) &=  \int \!\! \frac{d^3 p}{(2\pi)^3} \int \!\! \frac{d^3 q}{(2\pi)^3} \,
24 \, F_2^{(\rm s)}(\vkp,\vk-\vkp) F_4^{(\rm s)}(-\vkp,\vkp-\vk, \vp,-\vp) 
P_{11}(q) P_{11}(p) P_{11}(|\vk-\vkp|)\ , \\
\nn
P_{33}^{\rm (I)}(k) &= \int \!\! \frac{d^3 p}{(2\pi)^3} \int \!\! \frac{d^3 q}{(2\pi)^3} \,
 9 F_3^{(\rm s)}(\vk,\vq,-\vq) F_3^{(\rm s)}(-\vk,\vp,-\vp) P_{11}(k) P_{11}(q) P_{11}(p)\ , \\
 \nn
P_{33}^{\rm (II)}(k) &= \int \!\! \frac{d^3 p}{(2\pi)^3} \int \!\! \frac{d^3 q}{(2\pi)^3} \,
 6 F_3^{(\rm s)}(\vkp,\vp,\vk-\vkp-\vp) F_3^{(\rm s)}(-\vkp,-\vp,-\vk+\vkp+\vp) \\
&\qquad\qquad\qquad\qquad\qquad\qquad \times P_{11}(q) P_{11}(p) P_{11}(|\vk-\vkp-\vp|)\ ,
\end{align}
where the kernels $F_{3,4,5}^{(\rm s)}$ can be calculated from the recurrence relations found, e.g., in~\cite{Bernardeau:2001qr}, and symmetrizing over all arguments. We repeat these relations here for convenience:
\bea
\nn
F_n(\vq_1,\dots,\vq_n) &=& \sum_{m=1}^{n-1} \frac{G_m(\vq_1,\dots,\vq_m)}{(2n+3)(n-1)}
\left[ (2n+1) \frac{\vk\cdot\vk_1}{k_1^2} F_{n-m}(\vq_{m+1},\dots,\vq_n) \right. \\
\nn
&&\qquad\qquad\qquad\qquad\qquad \left. +  \frac{k^2 (\vk_1\cdot\vk_2)}{k_1^2 k_2^2}
G_{n-m}(\vq_{m+1},\dots,\vq_n) \right], \\
\nn
G_n(\vq_1,\dots,\vq_n) &=& \sum_{m=1}^{n-1} \frac{G_m(\vq_1,\dots,\vq_m)}{(2n+3)(n-1)}
\left[ 3 \frac{\vk\cdot\vk_1}{k_1^2} F_{n-m}(\vq_{m+1},\dots,\vq_n) \right. \\
&&\qquad\qquad\qquad\qquad\qquad \left. + n \frac{k^2 (\vk_1\cdot\vk_2)}{k_1^2 k_2^2}
G_{n-m}(\vq_{m+1},\dots,\vq_n) \right]\ ,
\eea
where $\vk_1=\vq_1+\cdots+\vq_m$, $\vk_2=\vq_{m+1}+\cdots+\vq_n$, $\vk=\vk_1+\vk_2$, and $F_n=G_n=1$.
From now on, we define the integrands $p_{51}$, $p_{42}$, $p_{33}^{\rm (I),(II)}$ by
\beq
P_{ij}(k) = \int \frac{d^3q}{(2\pi)^3} \int \frac{d^3p}{(2\pi)^3} \; p_{ij}(\vk,\vq,\vp)\ ,
\eeq
and we also use the notation $\mu=(\vkdq)/(kq)$ and $\nu=(\vk\cdot\vp)/(kp)$. 

We want to write all the 2-loop integrals as only one integral where the integrand is a sum of terms, each one with its leading divergence located only at $\vq\to 0$ \& $\vp\to 0$. In addition, each diagram will have subleading divergences occurring when only one combination of internal momenta tends to zero (for example, $\vq\to 0$ \& $\vp$ fixed, or $\vp\to 0$ \& $\vq$ fixed), and we would like to map all such divergences to $\vq\to 0$ \& $\vp$ fixed. Here, we examine each of the four integrands separately:

\begin{enumerate}
\item $P_{51}$ only has a single leading IR divergence, at $q\to 0$ \& $p\to 0$, so it already satisfies our first condition. It also has subleading divergences when either $\vq$ or $\vp$ approach zero while the other one remains fixed. Since $p_{51}(\vk,\vq,\vp)$ is symmetric in $\vq$ and $\vp$, we can perform the following manipulations to move the divergence at $\vp\to 0$ to $\vq\to 0$:
\begin{align}
\nn
P_{51}(k) &= \int \frac{d^3q}{(2\pi)^3} \int \frac{d^3p}{(2\pi)^3} \; p_{51}(\vk,\vq,\vp) 
\left[ \Theta(p-q) + \Theta(q-p)  \right]\\
&= \int \frac{d^3q}{(2\pi)^3} \int \frac{d^3p}{(2\pi)^3} \; 2\;p_{51}(\vk,\vq,\vp) \, \Theta(p-q)\ .
\label{eq:p51rearrange}
\end{align}

It is also possible to find the explicit forms of the various divergences. The leading divergence takes the following form (omitting a factor of $(2\pi)^{-5}$ and including the factor $q^2 p^2$ from the integration measure, since it affects the degree of divergence of the integral):
\beq
\label{eq:p51lead}
q^2 p^2 \, 2p_{51}(\vk,\vq,\vp) \underset{\vq\to 0, \, \vp\to 0}{\sim}  \frac{1}{2} k^{4+n} \mu^2\nu^2 q^n p^n \ ,
\eeq
while both the leading and subleading behavior is captured\footnote{Deriving this expression analytically is not straightforward given the large number of terms in $p_{51}$.  One can instead check that they agree numerically.} in the following expression:
\beq
q^2 p^2 \, 2p_{51}(\vk,\vq,\vp) {\underset{\vq\to 0}{\sim}}
\frac{k^{4+n} \mu^2(21k^4\nu^2+p^4[-10+59\nu^2-28\nu^4]-2k^2p^2[5-22\nu^2+38\nu^4])}
{42([k^2+p^2]^2-4k^2p^2\nu^2)} q^n p^n \ ,
\label{eq:p51sub}
\eeq
{with further subleading divergences discussed later.} We will find that both the leading and subleading divergences will cancel when summed with the other diagrams.
\item The leading divergences of $P_{42}$ are at $p\to 0$ \& $q\to 0$, and $p\to 0$ \& $\vkp\to\vk$, so we must manipulate the integrand to re-map the latter divergence to $p\to 0$ \& $q\to 0$. The algebra works the same as for $P_{22}$: split the $q$ integral into $q<|\vk-\vkp|$ and $q>|\vk-\vkp|$ pieces, and use the substitution $\vec{\tilde{q}}=\vk-\vkp$ on the second piece. We end up with
\beq
P_{42}(k) =   \int \frac{d^3q}{(2\pi)^3} \int \frac{d^3p}{(2\pi)^3} \; 2\; p_{42}(\vk,\vq,\vp) \,
\Theta(|\vk-\vkp|-q)\ ,
\eeq
which, however, has a subleading divergence at $\vp\to 0$ in addition to the one at $\vq\to 0$. To remedy this, we first symmetrize the integrand in $\vq$ and $\vp$ (recall from Eq.~(\ref{eq:2loopints}) that $p_{42}(\vk,\vq,\vp)$ is {\it not} symmetric in $\vq$ and $\vp$), then perform the same steps as in Eq.~(\ref{eq:p51rearrange}) to restrict the domain of integration to $p>q$. The result is
\beq
P_{42}(k) = \int \frac{d^3q}{(2\pi)^3} \int \frac{d^3p}{(2\pi)^3} \;
2 \left[ p_{42}(\vk,\vq,\vp) \, \Theta(|\vk-\vkp|-q)
+ (\vq \leftrightarrow \vp) \right] 
\Theta(p-q)\ .
\eeq
The leading divergence is 
\beq
\label{eq:p42lead}
q^2 p^2 \, 4p_{42}(\vk,\vq,\vp) \underset{\vq\to 0, \, \vp\to 0}{\sim}  - 2k^{4+n} \mu^2\nu^2 q^n p^n \ .
\eeq
The subleading divergences for $p_{42}(\vk,\vq,\vp)$ and $p_{42}(\vk,\vp,\vq)$ must be treated separately because they contribute over different domains in $p$.  One finds
\bea
\nonumber
&& q^2 p^2 2p_{42}(\vk,\vq,\vp)  \underset{\vq\to 0}{\sim}  \\ \nonumber
\label{eq:p42sub}
 &&\quad - \frac{k^{4+n} \mu^2(21k^4\nu^2+p^4[-10+59\nu^2-28\nu^4]-2k^2p^2[5-22\nu^2+38\nu^4])}
{21([k^2+p^2]^2-4k^2p^2\nu^2)} q^n p^n\\
&& q^2 p^2 2p_{42}(\vk,\vp,\vq)  \underset{\vq\to 0}{\sim} 
- \frac{k^{6+n}\mu^2 (7k\nu+p[3-10\nu^2])^2}
{49(k^2+p^2-2kp\nu)^{2+n/2}} q^n p^n\ .
\eea

\item $P_{33}^{(\rm I)}$ has its leading divergence at $p\to 0$ \& $q\to 0$, and subleading divergences at $\vq\to 0$ and $\vp\to 0$. It can therefore be rewritten in the same way as $P_{51}$:
\beq
P_{33}^{\rm (I)}(k) = \int \frac{d^3q}{(2\pi)^3} \int \frac{d^3p}{(2\pi)^3} \; 2\;p_{33}^{\rm (I)}(\vk,\vq,\vp) \, \Theta(p-q)\ .
\label{eq:p51rearrange}
\eeq
Indeed, it diverges in the same way as $P_{51}$, with leading piece
\beq
\label{eq:p33alead}
q^2 p^2 \, 2p_{33}^{\rm (I)}(\vk,\vq,\vp) \underset{\vq\to 0, \, \vp\to 0}{\sim}  \frac{1}{2} k^{4+n} \mu^2\nu^2 q^n p^n \ .
\eeq
and subleading expression
\begin{align}
q^2 p^2 \, 2p_{33}^{\rm (I)}(\vk,\vq,\vp) &\underset{\vq\to 0}{\sim}
\frac{k^{4+n} \mu^2(21k^4\nu^2+p^4[-10+59\nu^2-28\nu^4]-2k^2p^2[5-22\nu^2+38\nu^4])}
{42([k^2+p^2]^2-4k^2p^2\nu^2)} q^n p^n \ .
\label{eq:p33asub}
\end{align}

\item $P_{33}^{(\rm II)}$ has three leading IR divergences, at $q \to 0$ \& $p \to 0$, $\vq \to \vk$ \& $p \to 0$ and $\vq \to 0$ \& $p \to \vk$. There are also three subleading divergences: $\vq\to 0$ \& $\vp$ fixed, $\vp\to 0$ \& $\vq$ fixed, and $\vp\to \vk-\vq$. Thankfully, these can all be handled in a systematic way. First, we rewrite the integral to be symmetric in $\vq$, $\vp$, and a third internal momentum $\vl$:
\bea
\nn
P_{33}^{\rm (II)}(k) &=& \int \!\! \frac{d^3 p}{(2\pi)^3} \int \!\! \frac{d^3 q}{(2\pi)^3} \,
 6 F_3^{(\rm s)}(\vkp,\vp,\vk-\vkp-\vp) F_3^{(\rm s)}(-\vkp,-\vp,-\vk+\vkp+\vp) \\
 \nn
 &&\qquad\qquad\qquad\qquad \times P_{11}(q) P_{11}(p) P_{11}(|\vk-\vkp-\vp|) \\
 \nn
 &=& \int \!\! \frac{d^3 p}{(2\pi)^3} \int \!\! \frac{d^3 q}{(2\pi)^3} \int \!\! \frac{d^3 \ell}{(2\pi)^3}
 6 F_3^{(\rm s)}(\vkp,\vp,\vl) F_3^{(\rm s)}(-\vkp,-\vp,-\vl) \\
  &&\qquad\qquad\qquad\qquad \times P_{11}(q) P_{11}(p) P_{11}(\ell) \,
  \delta_{\rm D}(\vk-\vq-\vp-\vl)\ .
\eea
Written this way, the leading divergences occur when two of $\vq$, $\vp$, and $\vl$ go to zero, and the subleading divergences occur when only one momentum does. We can split the region of integration using the following sum of step functions:
\beq
\Theta(p-q) \left[ \Theta(\ell-p) + \Theta(p-\ell) \Theta(q-\ell) + \Theta(p-\ell) \Theta(\ell-q) \right]
+ (\vq\leftrightarrow\vp)\ .
\eeq
Since the integrand is symmetric in all three momenta, we are free to relabel the momenta in each term; after doing so, we find that each one is equivalent, so that we can write the sum as $6\Theta(p-q)\Theta(\ell-p)$, and the integral as
\bea
\nn
P_{33}^{\rm (II)}(k) &=& \int \!\! \frac{d^3 p}{(2\pi)^3} \int \!\! \frac{d^3 q}{(2\pi)^3} \int \!\! \frac{d^3 \ell}{(2\pi)^3}
 6 F_3^{(\rm s)}(\vkp,\vp,\vl) F_3^{(\rm s)}(-\vkp,-\vp,-\vl) \\
  &&\qquad\qquad \times P_{11}(q) P_{11}(p) P_{11}(\ell)
  \delta_{\rm D}(\vk-\vq-\vp-\vl) \times 6 \, \Theta(p-q) \Theta(\ell-p)\ .
\eea
This ensures that the leading and subleading divergences are all mapped to the same locations{: ($\vq\to0$ \& $\vp\to0$ or $\vq\to 0$ \& $\vp$ fixed)}, as desired. Upon evaluating the delta function, we are left with the original integrand $p_{33}^{\rm (II)}$ times a pair of step functions:
\beq
P_{33}^{\rm (II)}(k) =  \int \!\! \frac{d^3 p}{(2\pi)^3} \int \!\! \frac{d^3 q}{(2\pi)^3}
\;6\; p_{33}^{\rm (II)}(\vk,\vq,\vp) \, \Theta(p-q) \Theta(|\vk-\vq-\vp|-p)\ .
\eeq
Written like this, the integrand's leading IR behavior is
\beq
\label{eq:p33blead}
q^2 p^2 \, 6p_{33}^{\rm (II)}(\vk,\vq,\vp)
\underset{\vq\to 0, \, \vp\to 0}{\sim}   k^{4+n} \mu^2\nu^2 q^n p^n \ ,
\eeq
while the subleading behavior is
\beq
\label{eq:p33bsub}
6q^2 p^2 \, p_{33}^{\rm (II)}(\vk,\vq,\vp) \underset{\vq\to 0}{\sim}
\frac{k^{6+n}\mu^2 (7k\nu+p[3-10\nu^2])^2}
{49(k^2+p^2-2kp\nu)^{2+n/2}} q^n p^n\ .
\eeq
\end{enumerate}

In summary, we have rewritten the 2-loop integrand in the following way:
\bea
\nn
\tilde{p}_\text{2-loop}(\vk,\vq,\vp) &=&  
\left\{
2p_{51}(\vk,\vq,\vp) + 2p_{33}^{\rm (I)}(\vk,\vq,\vp) \right. \\
\nn
&&\qquad \left. + 2\left[ p_{42}(\vk,\vq,\vp) \, \Theta(|\vk-\vkp|-q) +
p_{42}(\vk,\vp,\vq) \, \Theta(|\vk-\vp|-p) \right]  \right. \\
&&\qquad \left.
+ 6p_{33}^{\rm (II)}(\vk,\vq,\vp) \, \Theta(|\vk-\vq-\vp|-p)
\right\} \Theta(p-q)\ ,
\label{eq:p2loopshort}
\eea
or, rewritten in terms of SPT kernels and linear power spectra,
\bea
\nn
\tilde{p}_\text{2-loop}(\vk,\vq,\vp) &=& 
\left\{ 60 F_5^{(\rm s)}(\vk,\vkp,-\vkp,\vp,-\vp) P_{11}(k) P_{11}(q) P_{11}(p) \right. \\
\nn
&&\left. + 18 F_3^{(\rm s)}(\vk,\vkp,-\vkp) F_3^{(\rm s)}(-\vk,\vp,-\vp)
	P_{11}(k) P_{11}(q) P_{11}(p) \right. \\
\nn
&&\left. + 48F_2^{(\rm s)}(\vkp,\vk-\vkp) F_4^{(\rm s)}(-\vkp,\vkp-\vk, \vp,-\vp) 
P_{11}(q) P_{11}(p) P_{11}(|\vk-\vkp|) \, \Theta(|\vk-\vkp|-q) \right. \\
\nn
&&\left. + 48F_2^{(\rm s)}(\vp,\vk-\vp) F_4^{(\rm s)}(-\vp,\vp-\vk, \vq,-\vq) 
P_{11}(q) P_{11}(p) P_{11}(|\vk-\vp|) \, \Theta(|\vk-\vp|-p) \right. \\
\nn
&&\left. + 36 F_3^{(\rm s)}(\vkp,\vp,\vk-\vkp-\vp) F_3^{(\rm s)}(-\vkp,-\vp,-\vk+\vkp+\vp) \right. \\
&&\qquad \left. \times P_{11}(q) P_{11}(p) P_{11}(|\vk-\vkp-\vp|) \, \Theta(|\vk-\vq-\vp|-p)
\right\} \Theta(p-q)\ .
\label{eq:p2looplong}
\eea
By summing Eqs.~(\ref{eq:p51lead}), (\ref{eq:p42lead}), (\ref{eq:p33alead}), and (\ref{eq:p33blead}), we see that the leading IR divergences have cancelled. In addition, by examining Eqs.~(\ref{eq:p51sub}), (\ref{eq:p42sub}), (\ref{eq:p33asub}), and (\ref{eq:p33bsub}), we see that the subleading divergences with asymptotic behavior $q^n$ as $\q\to 0\; \&\;\vp$ fixed have also cancelled. In particular, we can notice that in the limit of $q\ll k,p$, the structure of the $\Theta$-functions forces
two subsets of the contributions to cancel independently among themselves. In practice, we have
\bea
&& \tilde{p}_\text{2-loop}(\vk,\vq,\vp) \underset{q\ll k,\;q\ll p}{\simeq} \ \Theta(p-q) \ \times  \\ \nonumber
&&\left\{ P_{11}(k) P_{11}(q) P_{11}(p) \ \times\right. \\ \nonumber
&& \left.\left(60 F_5^{(\rm s)}(\vk,\vkp,-\vkp,\vp,-\vp) + 18 F_3^{(\rm s)}(\vk,\vkp,-\vkp) F_3^{(\rm s)}(-\vk,\vp,-\vp) + 48F_2^{(\rm s)}(\vkp,\vk-\vkp) F_4^{(\rm s)}(-\vkp,\vkp-\vk, \vp,-\vp) 
\right) \right. \\
\nn
&&\left. +  \Theta(|\vk-\vp|-p) \ P_{11}(q) P_{11}(p) P_{11}(|\vk-\vp|)\ \times \right. \\ \nonumber 
&&\left. \left(48F_2^{(\rm s)}(\vp,\vk-\vp) F_4^{(\rm s)}(-\vp,\vp-\vk, \vq,-\vq)  + 36 F_3^{(\rm s)}(\vkp,\vp,\vk-\vkp-\vp) F_3^{(\rm s)}(-\vkp,-\vp,-\vk+\vkp+\vp)  \right)  \right\} \ .
\label{eq:p2looplonglow}
\eea
Since for $|\vk-\vp|-p<0$ the terms in the fourth and fifth line vanish, this means that the IR divergences from the terms in the third lines cancel independently among themselves. This in turn implies that the terms in the fifth line cancel independently among themselves.

For $n>-2$, there are no remaining IR divergences, since terms scaling like $q^{n+1}$ or higher will converge when integrated. For $n<-2$, terms with $\sim$$q^{n+1}$ asymptotic behavior will result in divergences. However, these terms all involve odd powers of $\mu$ or $\nu$ (recall that these represent $\hat{k}\cdot\hat{q}$ and $\hat{k}\cdot\hat{p}$ respectively), which vanish when the angular integrals are performed. Alternatively, these terms vanish when the integrand is symmetrized in $\vq$ \& $-\vq$ and $\vp$ \& $-\vq$, so this is the final step we must take to ensure cancellation of divergences in the integrand itself for $n>-3$.

Therefore, the following form\footnote{
It should be noticed that the analytic string-length of the integrand can increase when rewritten in this form, but numerical evaluation, which is what is being addressed here, need not concern itself with such a proliferation in terms.  One simply defines functions as we do in eqns.~[\ref{eq:p2looplong}-\ref{symmIRSafe}], which are called with different numerical arguments. As the range of the integrand shrinks to match such duplication, no additional calculations are {\it a priori} preformed.  Moreover, adaptive monte-carlo series converge more readily under friendly integrals, making very manifest the computational benefits. } of $P_\text{2-loop}$ is free of IR divergences {\it before} integration:
\bea \nn
\label{symmIRSafe}
P_{\text{2-loop IR-safe}}(k) &=&  \int \!\! \frac{d^3 p}{(2\pi)^3} \int \!\! \frac{d^3 q}{(2\pi)^3}
\frac{1}{4}
\left[ \tilde{p}_\text{2-loop}(\vk,\vq,\vp) + \tilde{p}_\text{2-loop}(\vk,-\vq,\vp) \right. \\
&&\qquad\qquad\qquad\qquad
\left. + \tilde{p}_\text{2-loop}(\vk,\vq,-\vp) + \tilde{p}_\text{2-loop}(\vk,-\vq,-\vp) \right]\ ,
\eea
where $\tilde{p}_\text{2-loop}(\vk,\vq,\vp)$ is given in Eqs.~(\ref{eq:p2loopshort}) and~(\ref{eq:p2looplong}).

As $\q\to 0$ \& $\p\to 0$, the remaining integrand scales with momenta like $q^{n+2}p^{n+2}$, while for $q\to 0$ with $\vp$ fixed, it scales like $q^{n+2}$. As previously discussed, this makes an accurate numerical calculation of $P_\text{2-loop}$ much easier, as the delicate cancellation of large values of the integrand at different locations in momentum space is no longer necessary.

\subsection{Higher Loops}

IR divergences will cancel at any loop order for equal time correlators~\cite{Scoccimarro:1995if}. It is to be expected that the procedure we have outlined at one and two loops to define a global IR-safe integrand should work relatively unchanged at higher loops. The main steps of the procedure should be the following. As naively written, inside $L$-loop integrands the linear power spectra will be functions of linear combinations of $L$-loop momenta and external momenta.  The first step is to explicitly introduce one  momentum-conserving Dirac $\delta_D$-function so that all linear power spectra that appear in the integrand are functions only of individually labeled momenta $\{\vec q_i\}$'s, and not of combinations of $\vec q_i$ among themselves or with $\vec k$~\footnote{Why is only one additional Dirac $\delta_D$-function necessary no matter how high you go in loop order? Because each diagram is a sewing of $L+1$ primordial fluctuations in two tree diagrams, so only one ``cut" momenta ever need be written as a linear combination of the others, and it is that one that we resolve by introducing a Dirac $\delta_D$-function.}. Then, one symmetrizes
the integrand with respect to permutations of the subset $\{\vec q_j\}$ of the $\{\vec q_i\}$ that are involved in the Dirac $\delta_D$-function, and also with respect to all parity transformations acting
on the various $\{\vec q_j\}$'s. By inserting proper $\Theta$-functions and symmetry factors, one $q$-orders the various integrations over the moduli of the $\{\vec q_j\}$'s (similarly to the time-ordering procedure familiar in Quantum Field Theory). In this way one ensures that only one of the $\{\vec q_j\}$ momenta is allowed to go to zero. At this point, one symmetrizes with respect to all permutations including the remaining $\{\vec q_i\}$'s and all parities acting on those, and orders the resulting symmetric expression with $\Theta$-functions and symmetry factors so that only one momenta out of the full set of the $\{\vec q_i\}$'s is allowed to go to zero. As the last step, one evaluates the introduced Dirac $\delta_D$-function, so that the dimensionality of the final integral is no worse than when one started.

 \section{Application to Scaling Universes}
 
\subsection{Power Counting and Perturbative Expansion}

Let us apply our IR-safe integrand to the computation of the matter power spectrum at 2-loops in the case of a scaling universe where the tree level power spectrum
\be
P_{11}(k)=\frac{1}{\knl^3}\left(\frac{k}{\knl}\right)^n\ .
\ee
We choose the scaling universes because simple dimensional analysis allows us to infer the $k$-dependence of the answer. We will specialise to the cases $n=-3/2$ and $n=-5/2$ when doing the numerical calculations.

The first point to notice is that loop integrals will diverge in the ultraviolet (UV), the details depending on the tilt $n$. This simply means that the divergent contribution needs to be cancelled by counterterms that make the answer finite and cutoff-independent. The EFTofLSS provides the counteterms that cancel all physically possible UV divergences.  This was explained and applied to the current $\Lambda$CDM universe at 1-loop in~\cite{Carrasco:2012cv}, and later at 1-loop for the scaling universes in~\cite{Pajer:2013jj}. 
 
Following~\cite{Goroff:1986ep}, if $L$ is the number of loops, $I$ the number of internal lines,  $V$ the number of vertexes touching loops,  the superficial degree of divergence $D$ of a diagram is given, in terms of the tilt $n$ of the linear power spectrum, by
\be
D\leq 3 L+n I -V\ .
\ee
The factor of $3 L$ comes from the integration measure, the factor of $n I$ comes from the momentum-scaling of the internal lines, while the factor of $-V$ comes from the UV properties of the integrand $F^{(\rm s)}$ that vanishes at least as $1/q$, with $q$ being one of the high-scale internal momenta. This decoupling property is stronger when only one of the internal momenta goes to infinity. In that case the contribution from a vertex goes as $1/q^2$, but this property does not hold generically when more than one momentum goes to infinity.

Let us call $D_g$ the superficial degree of divergence of the loop diagram $P_g$, where the label $g$ is given by two numbers, each one representing the perturbative order of the two contributing $\delta$'s in the diagram. In order to cancel the IR divergences, we need to consider all diagrams at the same time, but the diagrammatic splitting is useful for studying the UV part, as this dictates which kind of counterterm of the EFTofLSS will remove the divergence. At 1-loop we have
\be
D_{13}=n+1\ ,\quad D_{22}=2n+1
\ee
 We see that $P_{13}$ can diverge for $n\geq -1$, while $P_{22}$ does so for steeper power spectra with $n\geq -1/2$.
 At two loops, we have
 \be
D_{15}=2n+5\ ,\quad D_{24,33}=3n+4\ .
\ee
We see that $P_{15}$ can diverge for $n\geq -5/2$, while $P_{24}$ and $P_{33}$ will generically diverge for steeper power spectra with $n\geq -4/3$.

The superficial degree of divergence represents an upper bound on the degree of divergence of a diagram. For example, it does not even take into account constraints from rotational invariance. Even more powerfully, one can check if a potential divergence can or cannot be reabsorbed by one of the counterterms of the EFTofLSS to further constrain the structure of the divergent terms. For example,  for $n\geq -5/2$ and for $P_{51}$, one can realize that the dependence on $\knl$ is fixed to be $\knl^{-3 (3+n)}$,  then use the superficial degree of divergence and finally the fact that $P_{51}$ has one $P_{11}$ out of the loop-integral, to conclude that $P_{51}$, evaluated with a UV cutoff $\Lambda$, takes the following form
 \bea\label{eq:p51_div}
&&P_{51}= \frac{1}{(2\pi)^5} \left[ c^\Lambda_{2n+5}  \left(\frac{\Lambda}{\knl}\right)^{2n+5}\left(\frac{k}{\knl}\right)^{1}P_{11} \right. \\ 
&& \nonumber \qquad\qquad\qquad
 \left. + \, c^\Lambda_{2n+4} \left(\frac{\Lambda}{\knl}\right)^{2n+4}\left(\frac{k}{\knl}\right)^{2}P_{11}+{\text{subleading divergent terms}} \right],
\eea
% \bea\label{eq:p51_div}
%&& \!\!\!\!  \!\!\!\!  \!\!\!\! P_{51}=\\
%&& \!\!\!\!  \!\!\!\!  \!\!\!\! \nonumber \quad c^\Lambda_{2n+5}  \left(\frac{\Lambda}{\knl}\right)^{2n+5}\left(\frac{k}{\knl}\right)^{1}P_{11}+c^\Lambda_{2n+4} \left(\frac{\Lambda}{\knl}\right)^{2n+4}\left(\frac{k}{\knl}\right)^{2}P_{11}+{\text{subleading divergent terms}},
%\eea
where the factors of $(2\pi)$ count the phase-space suppression of higher loop integrals, and the $c_i$ are order one numbers. If this answer were correct, there would need to be a {\it local} counterterm in the EFTofLSS that goes as $k\times P_{11}\sim k \langle\delta_k^2\rangle$. Since the power spectrum is proportional to $P_{11}(k)$, the counterterm must be proportional to $\delta$. Such a counterterm should therefore have the form of $k\, \delta$, but this breaks rotational invariance or locality (which requires analyticity in $\vec k$) and so is absent from the EFTofLSS. The lowest derivative available counterterm is the so-called speed of sound term of the form $k^2\delta$, which indeed can reabsorb the subleading divergence~\footnote{The same identical reasoning would have allowed us to remove a putative divergent term of the form
\be
 \frac{c^\Lambda_{2n+6}}{(2\pi)^5}  \left(\frac{\Lambda}{\knl}\right)^{2n+6}P_{11} \ ,
\ee
without having to deal with the UV properties of the vertex functions. Indeed, this term can be trivially excluded
due to the absence of a proper counterterm in the EFTofLSS.}. We can therefore drop the most divergent term, and conclude that 
 \bea\label{eq:p51_div2} 
&& P_{51}= \frac{1}{(2\pi)^5} \left[ c^\Lambda_{2n+4} \left(\frac{\Lambda}{\knl}\right)^{2n+4}\left(\frac{k}{\knl}\right)^{2}P_{11}+{\text{subleading divergent terms}} \right] \ .
\eea

For $n=-3/2$, the 1-loop terms are finite, but the 2-loop terms can (and actually will) diverge as $\Lambda^1$. In addition, there could be in principle a subleading logarithmic divergence of the form
\be
P_{51,\;n=-3/2}\supset \frac{c^\Lambda_{\log}}{(2\pi)^5} \log \left(\frac{\Lambda}{k}\right)\; \left(\frac{k}{\knl}\right)^3 P_{11}\ ,
\ee
for which there is no available counterterm after using the rotational invariance of the counterterm. We therefore conclude that the logarithmic divergence is absent: $c^\Lambda_{\log}=0$.  One can arrive at the same conclusion starting directly from the loop integral by expanding in $k / q_{\rm loop}$, where $q_{\rm loop} \to \Lambda$ is any of the diverging loop momenta.  The first subleading term (after the leading divergence) must be of the form $f(q_{\rm loop}) {\vec q}_{\rm loop} \cdot {\vec k}/q_{\rm loop}^2$ which vanishes after performing the angular intergrals. We see that the EFTofLSS allows us to infer information about the outcome of the integral even before actually doing the calculation and beyond naive dimensional analysis. It is interesting to verify these claims in a numerical calculation as we do next.

Although the logarithmic term is forbidden, a finite term
\beq
P_{51,\;n=-3/2}\supset \frac{c_\text{finite 1}}{(2\pi)^5} \;  \left(\frac{k}{\knl}\right)^3 P_{11}\ 
\eeq
is  allowed with $c_\text{finite 1} \neq 0$, since it cannot forbidden on the grounds of available counter-terms.  Because the divergences will all be absorbed by the counterterms, the finite terms are the only physically meaningful contribution to the power spectrum. In this case there is no scale in the computation except for the external wavenumber $k$ (the overall dependence on $\knl$ is irrelevant as it simply enter linearly in the answer), for the loop integral to be finite it means that the integrand is peaked around momenta of order $q_{\rm loop} \sim k$. Using dimensional analysis we therefore conclude that 
\be
P_\text{L-loops-finite}\sim \frac{1}{\knl^3}\left(\frac{k}{\knl}\right)^{(3+n)L} \left(\frac{k}{\knl}\right)^n\ .
\ee
This has the same powers of $\knl$ as (\ref{eq:p51_div2}), but with no powers of $\Lambda$ in the numerator. Notice that this is only a small correction for $k\ll \knl$. As discussed and used in~\cite{Carrasco:2012cv}, the EFTofLSS provides a manifestly convergent perturbative expansion in $k/\knl$ {\it after} renormalization has been performed. At this point each loop contributes a finite hierarchically-smaller correction in $k/\knl\ll 1$. We also see that we are really interested in the finite contributions. This means that the numerical precision in evaluating the integrals should be rather high. We now proceed to the numerical evaluation of the 2-loop integrals.

\subsection{Numerical Evaluation and Renormalization}

After manifestly handling the IR cancellation in the IR-safe integrand, it should be possible to numerically extract subleading in $k$ finite parts from fitting to integrated data for various tilts without being confounded by spurious noise. We demonstrate this is indeed the case in this section, where we measure the actual coefficients of the series expansion in $(k/\knl)$. 

\subsubsection{$n=-\frac{3}{2}$ tilt}

Because of the UV divergence in the 2-loop integrals  for the case $n=-3/2$, we will need to evaluate the 2-loop integrals with a cutoff $\Lambda$ and cancel the $\Lambda$ dependence with the speed of sound counterterm. This is given by
\be\label{eq:2-loop-counter}
P_{\text{2-loop counterterm}}= \frac{c_{\text{counter}}^\Lambda}{(2\pi)^5} \left(\frac{\Lambda}{\knl}\right)\left(\frac{k}{\knl}\right)^2 P_{11}\ ,
\ee
where the coefficient $c^\Lambda_\text{counter}$ is carefully chosen to cancel the UV divergence in $P_{\text{2-loop}}$.

We will evaluate the integrals numerically with a finite $\Lambda$ and with $k\ll \knl$. We expect
 \bea\label{eq:p51_div3} 
&& P_{\text{2-loop}}= \frac{1}{(2\pi)^5}\left[c^\Lambda_{1} \left(\frac{\Lambda}{\knl}\right)^{1}\left(\frac{k}{\knl}\right)^{2}P_{11}\right.
+c^{{\text{finite}}}_1  \left(\frac{k}{\knl}\right)^{3}P_{11}\\ \nonumber &&
\left.\qquad\qquad\qquad \quad+ c^{1/\Lambda}_{1} \left(\frac{k}{\Lambda}\right)^{1}\left(\frac{k}{\knl}\right)^{3}P_{11}+{\text{subleading finite terms in }} \frac{k}{\Lambda}\right] \ . 
\eea
The factors of $2\pi$ count the phase-space suppression of higher loop contributions.
In the last line, we have added terms that depend on $\Lambda$ in a way that converges to zero as $\Lambda\to \infty$. These terms are negligible only in the limit $\Lambda\to \infty$. If one does the calculation at finite $\Lambda$, or one measures the coefficients of the EFTofLSS from $N$-body simulations, these terms might need to be included, depending on the value of $k/\Lambda$. These terms are potentially important for us, as we evaluate the loops with a finite cutoff.

\begin{figure}[t]
\begin{center}
\includegraphics[width=0.49\textwidth]{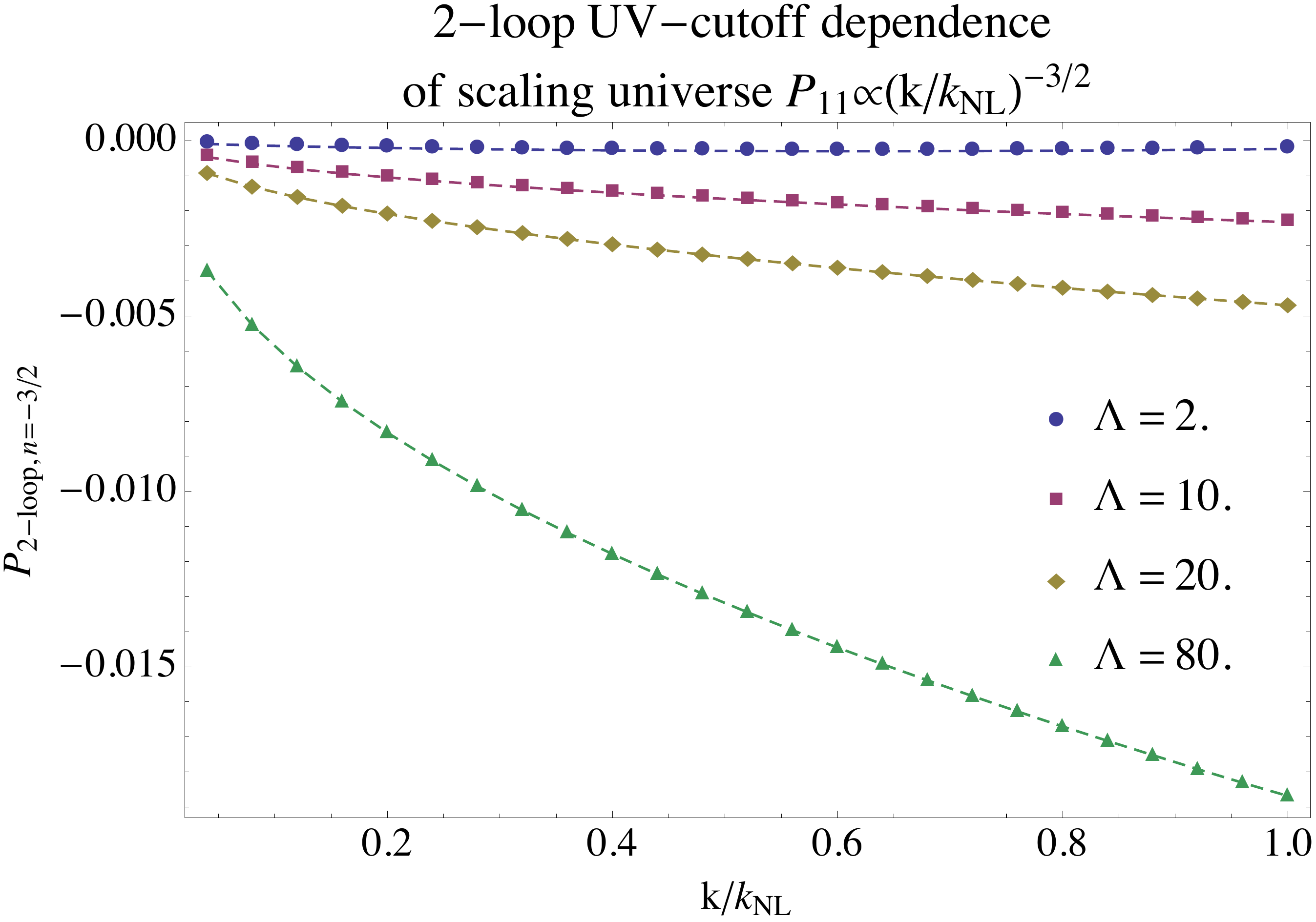}
\includegraphics[width=0.49\textwidth]{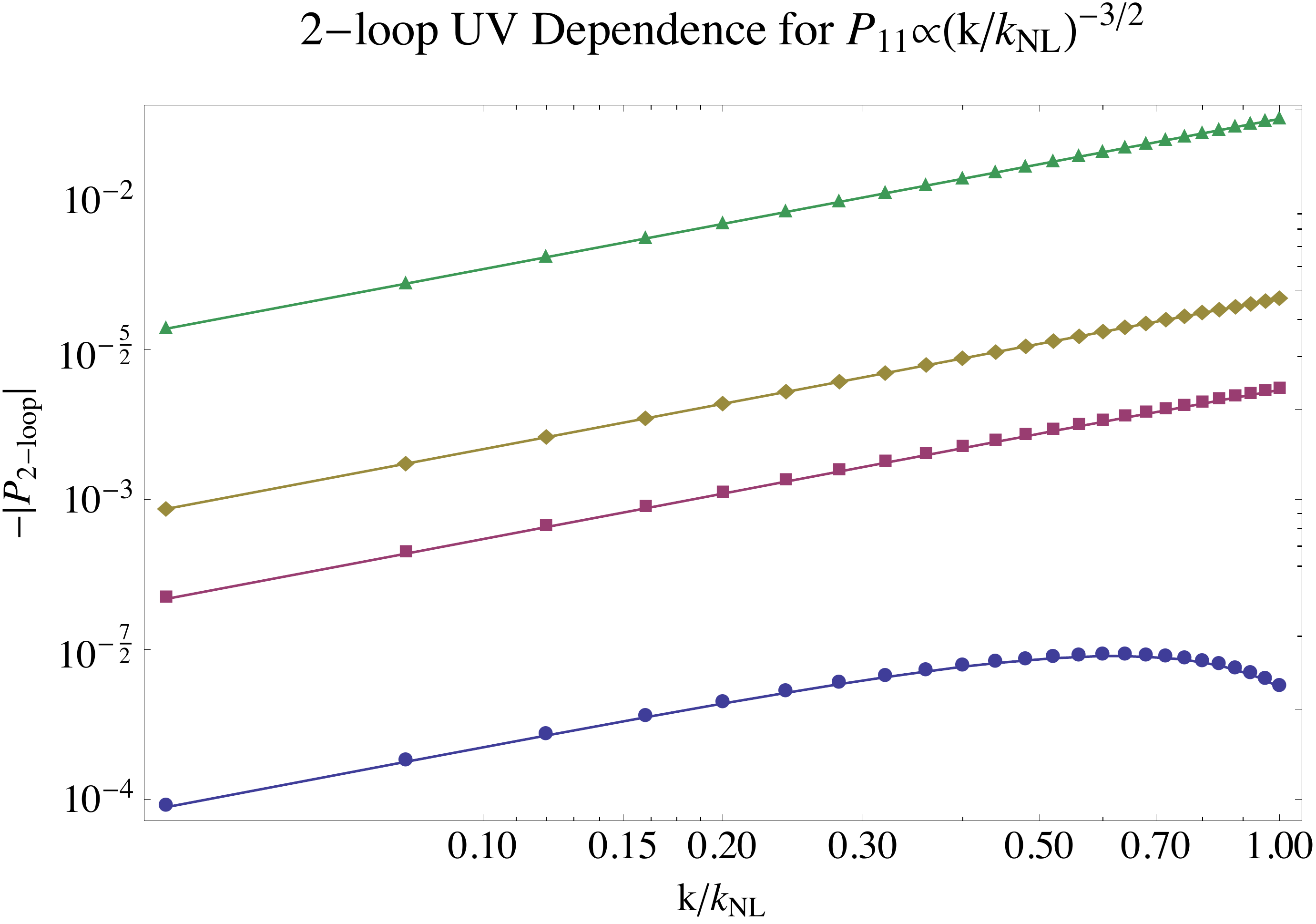}
\caption{\label{fig:variousLambdakmin} \small\it  Left: The 2-loop power spectrum $P_{\text{2-loop}}$ for various values of the UV cutoff $\Lambda$ and IR cutoff $\kmin$. We show $\Lambda=2\knl$ in blue, $10\,\knl$ in red, $20\,\knl$ in yellow, and $80\,\knl$ in green,  The result is $\kmin$ independent, as we demonstrate in detail in Fig.~\ref{fig:kmin_dependence}, so here we plot only results with $\kmin=5\times10^{-3}\knl$, as the results with $\kmin=5\times10^{-4}\knl$ are visually indistinguishable.  Right: same as above but on a logarithmic scale. When the UV cutoff is high enough so that finite terms are negligible, we can see that all the curves have the same slope, corresponding to $k^{1/2}$ dependence, and that they depend linearly on $\Lambda$. This agrees with what is expected from the divergent term in (\ref{eq:p51_div3}).}
\end{center}
\end{figure}

If we evaluate the integrals with a very high $\Lambda$, it is then numerically difficult to have enough precision to extract the subleading finite term. We proceed therefore in the following way. We first evaluate the 2-loop integral with a very high $\Lambda$, in such a way that we can neglect all the subleading terms and we can extract the coefficient $c^\Lambda_1$ by measuring low-$k$ scaling behavior. In Fig.~\ref{fig:variousLambdakmin} we plot the value of the 2-loop power spectrum for various values of $\Lambda$ and an IR cutoff $\kmin$. We see that for values of $\Lambda$ high enough so that the finite terms are negligible, the result goes as $\Lambda\, k^{1/2}$, as it should, with no additional $\log(\Lambda)$ dependence, as anticipated by considerations related to the absence of counterterms in the EFTofLSS.   This will be more clearly visible after we remove the linear divergence.  We also note here that the result is $\kmin$ independent, as we demonstrate in detail in Fig.~\ref{fig:kmin_dependence}. 

\begin{figure}[t]
\begin{center}
\includegraphics[width=0.8\textwidth]{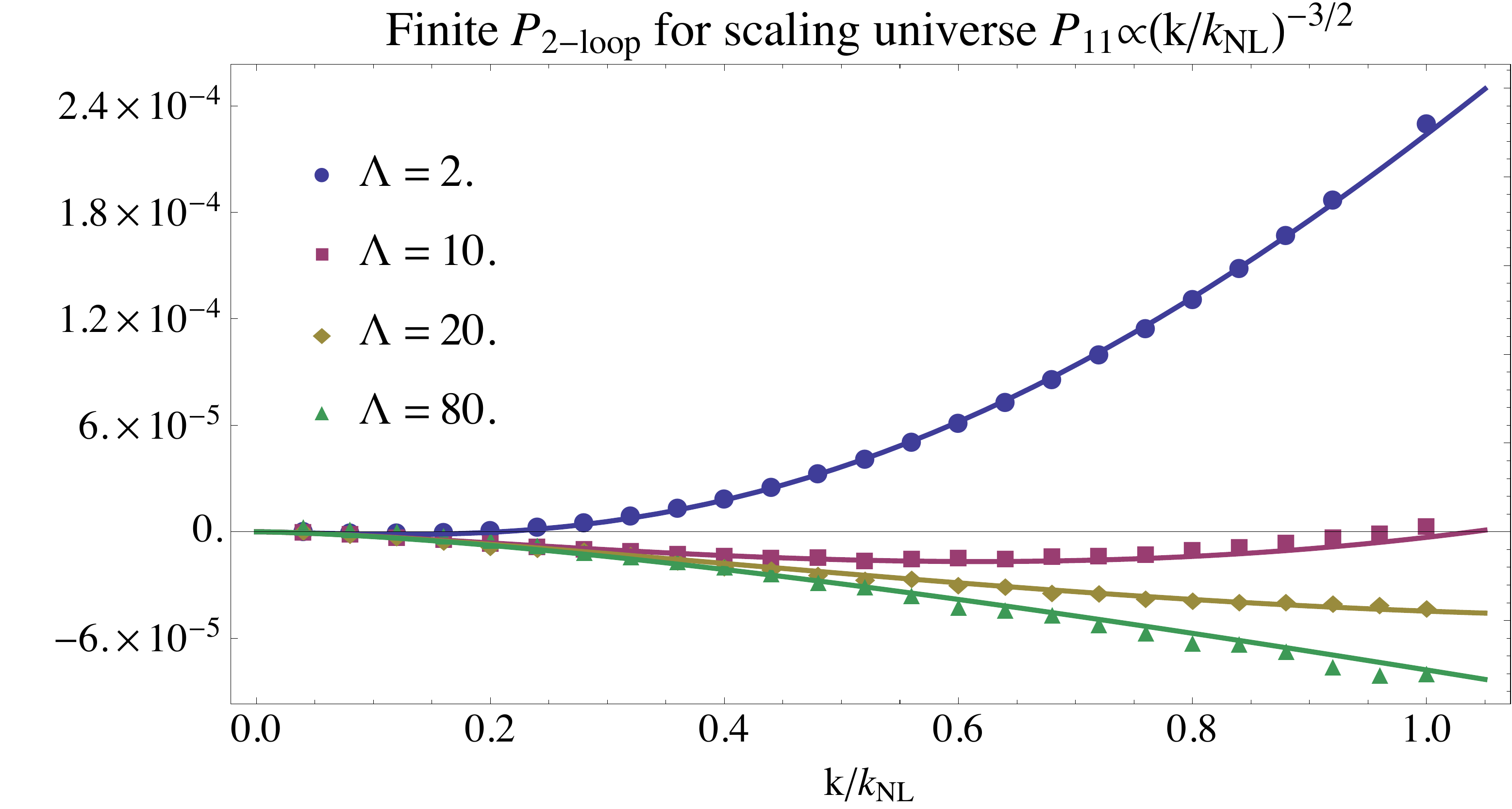}
\caption{\label{fig:variousLambdakmin2} \small\it The points represents the finite contributions to $P_{\text{2-loop}}$ for $\Lambda=2 \, \knl$ (blue) $\Lambda=10\,\knl$ (red), $\Lambda=20\,\knl$ (yellow) and $\Lambda=80\,\knl$ (green) obtained after the divergent part has been subtracted. As we go to low enough $k$'s, the three curves approach the same one, as indicated by the continuous fitting functions of Eq.~(\ref{eq:finite}), plotted as solid lines. As we move to higher $k$'s the $\Lambda$ dependence, strongest for the $\Lambda=2\,\knl$ case, becomes clearly visible.} 
\end{center}
\end{figure}

After we have measured $c^\Lambda_{1}$, we set 
\be
c_{\text{counter}}^\Lambda =-c^\Lambda_{1} + \delta c_{\text{counter}} \left(\frac{\knl}{\Lambda}\right) \ ,
\ee
where $\delta c_{\text{counter}}$ represents the finite ($\Lambda$-independent) contribution of the term in (\ref{eq:2-loop-counter}). As described above, we find $c_1^\Lambda$ by  evaluating $P_\text{2-loop}$ with a high cutoff $\Lambda=80\, \knl$, and we fit the result with a $k$ and $\Lambda$ dependence of the form
\be\frac{c^\Lambda_{1} }{(2\pi)^5}
\left(\frac{\Lambda}{\knl}\right)^{1}\left(\frac{k}{\knl}\right)^{2} P_{11}\ .
\ee
in the interval $0.04\, \knl \leq k\leq 0.16\; \knl$, so that the finite corrections, that become relatively smaller for $k\to 0$, are under control. We take $\kmin=0.005\, \knl$.   As demonstrated later in Fig.~\ref{fig:kmin_dependence}, decreasing kmin by order of magnitude is completely negligible.  
We numerically find
\be
c_1^\Lambda\simeq - 2.277\ ,
\ee
which is order one as expected.  We then evaluate the same $P_\text{2-loop}$ with four lower cutoffs $\Lambda=2 \, \knl,\ 10\, \knl$, $20\, \knl$ and $80\,\knl$. After subtracting the divergent term, we fit the remaining part with the following functional form
\be\label{eq:finite}
\frac{1}{(2\pi)^5}\left[ c^{{\text{finite}}}_1 
+c^{1/\Lambda}_1\left(\frac{k}{\Lambda}\right)  +c^{1/\Lambda}_2\left(\frac{k}{\Lambda}\right)^2 \right] \left(\frac{k}{\knl}\right)^{2(3+n)} P_{11}\ .
\ee
in the $k$ interval $0.04\, \knl \leq k\leq  \knl$. We numerically find
\be
 c^{{\text{finite}}}_1\simeq -0.874 \ , \quad c^{1/\Lambda}_1\simeq 9.005 \  , \quad c^{1/\Lambda}_2\simeq -5.745  \ , \quad n=-1.496\ .
\ee
The   coefficients are of order one, as expected. Importantly, the measured scaling $n$ agrees with the expected one $n=-3/2$. The fit comparison with the numerical data after the counterterm subtraction is shown in Fig.~\ref{fig:variousLambdakmin2}.  Of course by including terms higher order in ${k/\Lambda}$ one can improve the fit higher in $k/k_{NL}$, but the above is sufficient to demonstrate the convergence of the IR-behavior of the integrand.

\subsubsection{$n=-\frac{5}{2}$ tilt}

For a tilt of $n=-\frac{5}{2}$ the UV situation is even more straightforward as there is no dominant dependence on the UV cutoff, the integral being UV convergent.  This universe is a particularly stringent test of the IR behavior however. While it should be IR finite, there can and is measurable IR cutoff ($\kmin$) dependence, in a convergent way:
\begin{equation}
P_{\text{2-loop, n=-5/2}}= \frac{1}{(2\pi)^5}\frac{1}{\knl^3} \left[ c_1 \left( \frac{k}{\knl} \right)^{-3/2} + {\cal{O}}\left(\left(\frac{\kmin}{k} \right)^{1/2}\right)\right]\ .
\end{equation}
This is entirely consistent with the idea of IR divergence occurring with $n=-3$.  Consider for example the case of $n=-3+\epsilon$, with $0<\epsilon\ll1$. The dependence on $\kmin$ should be of order  $\left(\kmin/k \right)^\eps$, which is not a very subdominant correction for a given ratio $\kmin/k$ if $\epsilon$ is small enough. To this point, under various IR cutoffs, we find the $n=-5/2$ universe scales as follows:
%\begin{equation}
%\label{eqn:neg5o2Fit}
%P_{\text{2-loop, n=-5/2}}^{\text{fit}}= \frac{1}{(2\pi)^5}  \frac{1}{\knl^3} \left( \frac{k}{\knl} %\right)^{-3/2}  \left[499  -1557 \left( \frac{\kmin}{k} \right)^{1/2} + 1116  \left( \frac{\kmin}{k} \right) \right]\,.
%\end{equation}
\begin{equation}
\label{eqn:neg5o2Fit}
P_{\text{2-loop, n=-5/2}}^{\text{fit}}= \frac{1}{\knl^3} \left(\frac{k}{\knl} \right)^{-3/2}  \left[  0 .051  -0.159 \left( \frac{\kmin}{k} \right)^{1/2} + 0.114  \left( \frac{\kmin}{k} \right) \right]\,.
\end{equation}

\begin{figure}[t]
\begin{center}
\includegraphics[width=0.49\textwidth]{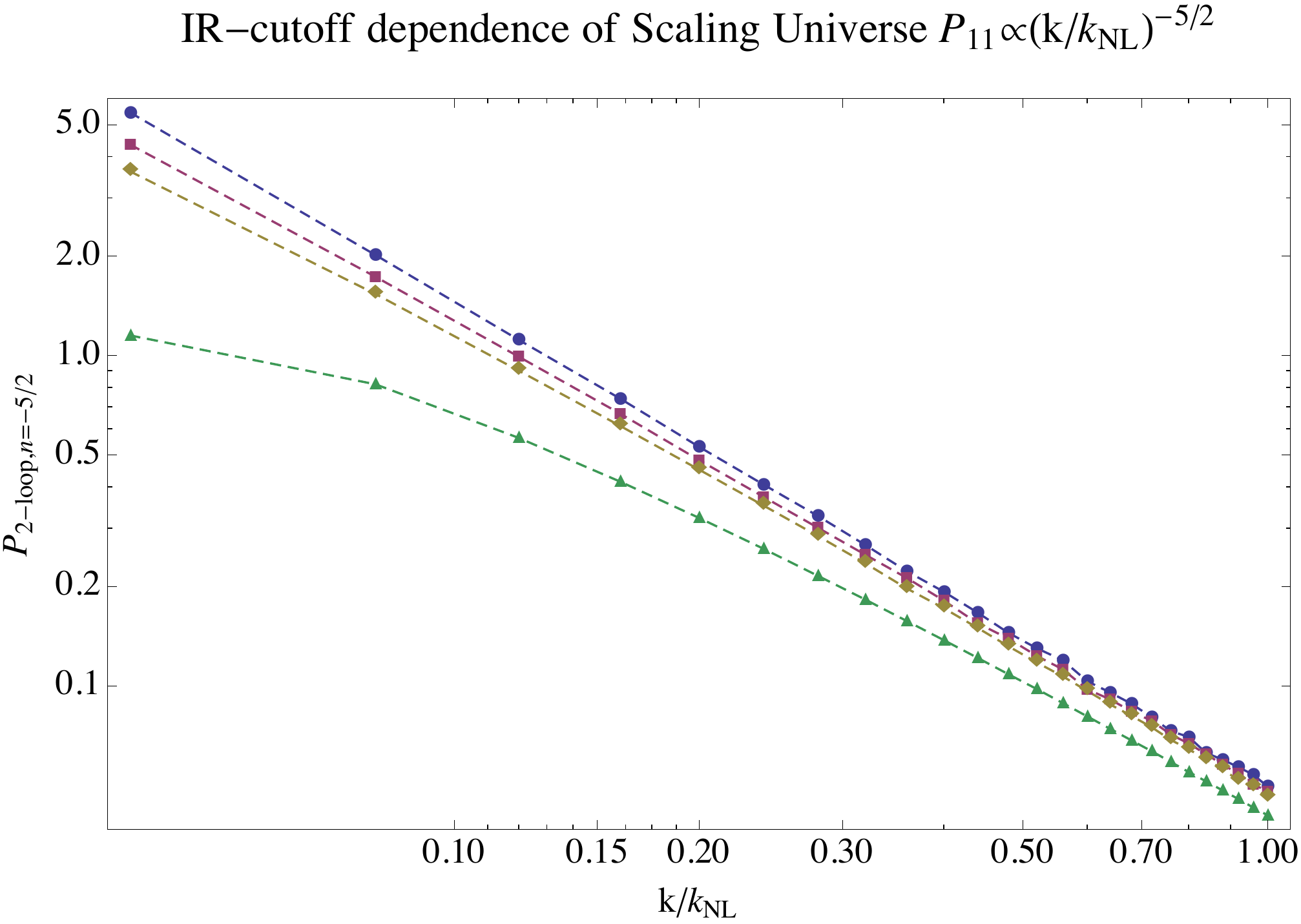}
\includegraphics[width=0.49\textwidth]{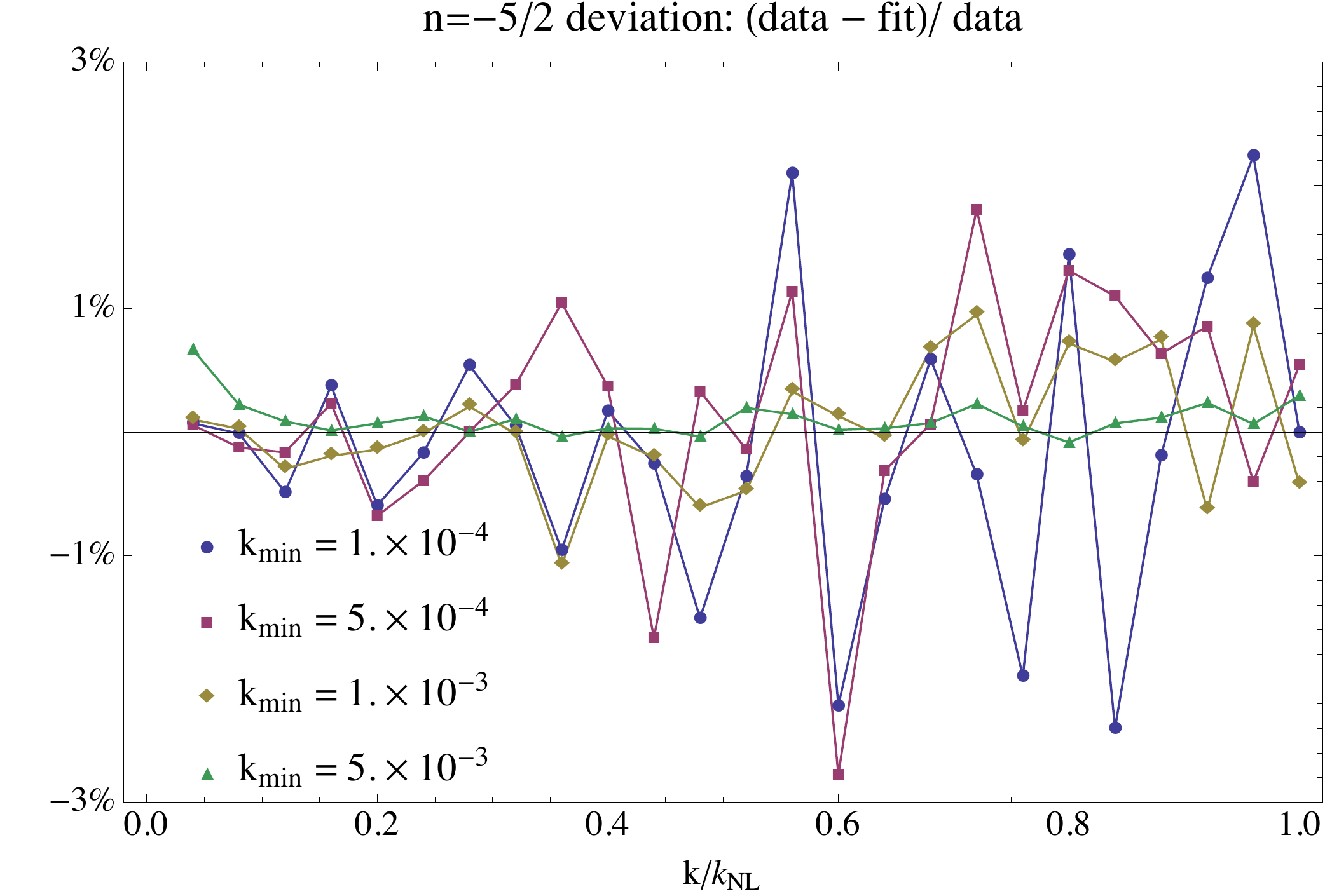}
\caption{\label{fig:neg5o2} \small\it Left: 2-loop power spectrum $P_{\text{2-loop}}$ for ideal scaling of tilt $n=-5/2$ for various values of the IR cutoff $\kmin$.  This universe is UV cutoff ($\Lambda$) independent.  Right: We demonstrate the functional form of the IR-dependence by comparing to the fit given in eq.~(\ref{eqn:neg5o2Fit}).   }
\end{center}
\end{figure}

We present the data, and comparison to the fit in Fig.~\ref{fig:neg5o2}. The agreement is very good.

\section{Summary}

In the upcoming era where large scale structure surveys will likely be the leading source of cosmological information, it is crucial to make reliable predictions for as many of the modes as possible in order to maximize the information that can be extracted. Since the phase space density of modes goes as $k^3$, with $k$ being the wavenumber of a mode, most of the information is contained in high wavenumbers, where non-linear corrections become important. In our current universe, there is a range of $k$-modes, roughly between $0.1\,h\,$Mpc$^{-1}$ and $1\,h\,$Mpc$^{-1}$, where the non-linearities of dark matter clustering are large enough to make it necessary for them to be taken into account, but also small enough to be amenable to an analytical treatment. Given the phase space density of modes, this interval in wavenumber is actually the most important for observations.

With the development of the Effective Field Theory of Large Scale Structures (EFTofLSS)~\cite{Baumann:2010tm,Carrasco:2012cv}, the theoretical framework where a manifestly convergent perturbative expansion in $k/\knl$ has been established, where $k$ is the wavenumber of the mode of interest, and $\knl$ is the one associated with the non-linear scale. By computing  higher order corrections, adding suitable counterterms order by order, and renormalizing the theory, more accurate predictions in $k/\knl$ can be made, approaching the non-linear scale $\knl$.

Given how much phase space is focussed at the UV and the absence of formally reliable analytical techniques,  the actual value of $\knl$ in our universe at order-one is still unclear. Having $\knl=0.5\,h\,$Mpc$^{-1}$ or $1\,h\,$Mpc$^{-1}$ makes a very large difference in the amount of available data we can use for cosmology. 

We suggest $\knl$ should be the scale at which the EFTofLSS stops converging.  To identify this, higher order calculations beyond the ones first studied in~\cite{Carrasco:2012cv} need to be computed. The next step is to compute the 2-loop correction. Beyond the standard numerical challenges,  one is faced with two additional subtleties. First, each of the standard two-loop diagrams has several IR divergences in different soft regions of the integrand that cancel only when all the diagrams are summed over.  This makes the numerical
evaluation much harder, as the largest contribution to each diagram is actually unphysical and will need to cancel out from the final answer. Second, the leading contribution of the IR-independent part is degenerate with the speed-of-sound counterterm in the EFTofLSS, and so is uncalculable. For this reason, the physically relevant 2-loop term is the next-to-leading IR-finite term.

In this paper we have presented a procedure which re-arranges all the four standard 2-loop diagrams into only one integral,  and  we manipulate the integrand in such a way that the only possible IR divergences appear when either both the internal momenta or only one of the two goes to zero. Since IR divergences are guaranteed to cancel from the final answer, our procedure enforces the cancellation at the level of the integrand. Now this `IR-safe global integrand' is IR integrable in any soft limit, and this happens both for the leading IR divergences where both the internal momenta go to zero, and also for the sub-leading ones where only one of the internal momenta go to zero. This makes the numerical integration much easier, as the integral is automatically of the size of the IR-independent part, without there being cancellation between regions of the integrand far-away in phase space.

After implementing our IR-safe integrand we have computed the 2-loop power spectrum in scaling universes with a linear power spectrum of the form $k^n$. They have the property of being particularly simple, so that the form of the final answer can be derived using dimensional analysis. At the same time these form the most strident tests from the IR-point of view, as the standard integrand does not go nicely to zero in the soft limit. In particular, we have explored  first the case $n=-3/2$, which has the property of being  UV divergent. We evaluate the integrals with a UV cutoff $\Lambda$, we renormalize the leading divergence, and we investigate the residual dependence on the IR cutoff $\kmin$ and verify that the final answer agrees with the behavior we predicted from dimensional grounds. Satisfactorily, we find that the dependence on $\kmin$ is subpercent, and that the residual analytic dependence of the 2-loop power spectra after renormalization perfectly matches our expectations. Subsequently, we have tested the case $n=-5/2$, which does not have UV divergences, so that the result of the integration is automatically finite and no renormalization is required with UV Divergent counterterms\footnote{Of course the finite counter terms contribute as some power of $(k/\knl)$ so if one were to match against simulations of tilt $-5/2$ scaling universes, finite counter terms should be included and appropriately renormalized.}, but that has a potentially larger IR divergence.  In this case as well, we have found that the answer depends correctly on $\kmin$, with corrections proportional to $(\kmin/k)^{1/2}$, which goes to zero as~$\kmin\to 0$. We have also tested our calculation against different numerical integration techniques (Montecarlo and Quasi-Montecarlo), finding that the result does not depend relevantly on the numerical technique used.

We have compared the performance of our technique for computing the 2-loop power spectrum with the standard techniques used in some publicly available codes. We have found that without the implementation of our IR-safe integrand, these code are unable to reliably compute the 2-loop power spectra for scaling universe, and probably also for our current universe.

In a forthcoming paper~\cite{nextpaper}, we will apply this integration technique to give the 2-loop prediction of the power spectrum from the EFTofLSS, where we will see that the EFTofLSS predicts the power spectrum with percent precision up to roughly $k\sim 0.5 \, h\,$Mpc$^{-1}$.

\section*{Acknowledgments}
We thank  Tobias Baldauf, Francis Bernardeau, Toni Riotto, Uro\v{s} Seljak, Zvonimir Vlah,   and Matias~Zaldarriaga for useful conversations.  We thank Francis Bernardeau and Atsushi Taruya for extensive help in running the RegPT code.  J.J.M.C. would like to thank thank Academic Technology Services at UCLA for computer support.  J.J.M.C.~is supported by the Stanford Institute for Theoretical Physics and the NSF grant no. PHY-0756174 and a 
grant from the  John Templeton Foundation.  S.F. is partially supported by the Natural Sciences and Engineering Research Council of Canada.   D.G.~is supported in part by the Stanford ITP and by the U.S. Department of Energy contract to SLAC no.\ DE-AC02-76SF00515.  L.S. is supported by DOE Early Career Award DE-FG02-12ER41854 and by NSF grant PHY-1068380.  The opinions expressed in this publication are those of the authors and do not necessarily reflect the views of the John Templeton Foundation.

\vspace{0.8cm}
{\bf Note Added:}  While this paper was being finalized Ref.~\cite{Blas:2013bpa} appeared, which may have some overlap.  Our preliminary results have already been presented at the Feb.~2013 workshop ``Dark Energy Phenomenology''~\cite{deWork}.

\appendix

\section*{Appendix}
\section{Numerical Checks}

\subsection{Independence of IR cutoff and of method of integration}

In this appendix we provide some details of the implementation and some explicit checks of numerical convergence.  

All numerical integrations of the global IR-safe integrand were performed using the CUBA library~\cite{CUBA} (v.~3.0).  The implementation of our IR-safe integrand used the recursive implementation of the kernels given in the Copter library~\cite{Carlson:2009it}, but combined as given above in eq.~(\ref{eq:p2looplong}). 

\begin{figure}[t]
\begin{center}
\includegraphics[width=0.8\textwidth]{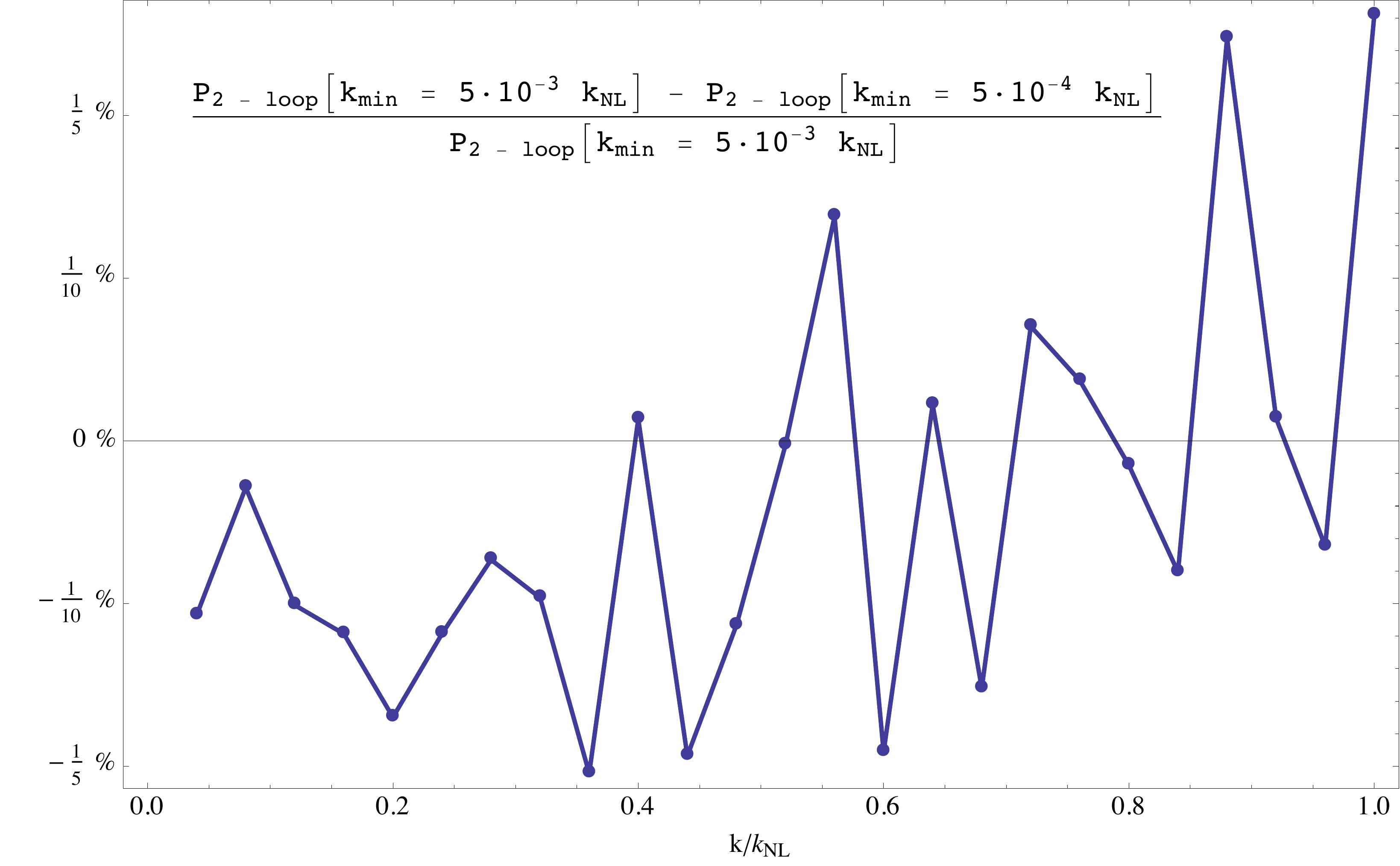}
\caption{\label{fig:kmin_dependence} \small\it Fractional difference between $P_{\text{2-loop},n=-3/2}$ for $\kmin=5 \times 10^{-3}\, \knl$ and $\kmin=5 \times 10^{-4}\, \knl$ for $\Lambda=10$. We see that for $k\lesssim 1$, the dependence is less than 1/3 of a percent.}
\end{center}
\end{figure}

First we demonstrate that our results are independent of the IR cutoff.  In Fig.~\ref{fig:kmin_dependence}, we plot the relative difference for $P_\text{2-loop}$ obtained for different values of $\kmin$. We see that over the range explored, the $k_{\rm min}$-dependence is at the less than percent level, becoming of order 0.3\% at $k\simeq \knl$.

\begin{figure}[t]
\begin{center}
\includegraphics[width=0.8\textwidth]{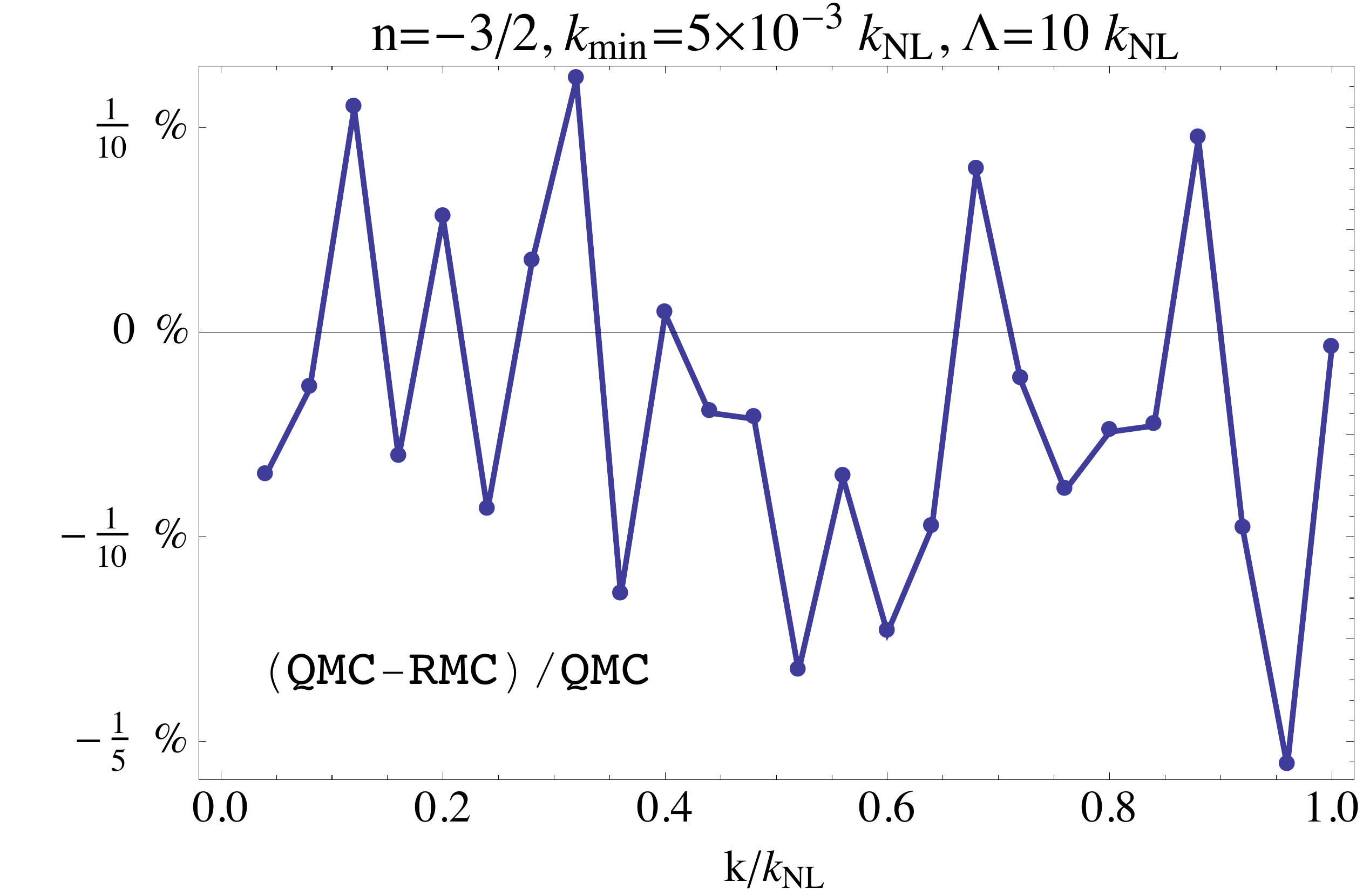}
\caption{\label{fig:integration_method} \small\it  Plot of the deviation between the ``quasi-Monte-Carlo" and "random Monte-Carlo" results from CUBA's Vegas integration algorithm, for
the IR safe integrand for scaling universe $n=-3/2$.  We see that it is sub-percent for the range considered. }
\end{center}
\end{figure}

We demonstrate our results are independent of the method of integration  in Fig.~\ref{fig:integration_method}.  We find that there is sub-percent relative difference in  $P_\text{2-loop}$ computed for $\Lambda=2$, using two methods, ``quasi-Monte-Carlo" and ``random Monte-Carlo," for the Vegas algorithm in CUBA.
Both were set to terminate when achieving either a relative (estimated) accuracy goal of $10^{-3}$, or reaching a maximum of $10^8$ integrand evaluations. 

\subsection{Comparison with available 2-loop results}

In this section we compare our results with other publicly available numerical tools and their associated 2-loop integrands.
We do this to put into evidence the importance of performing the numerical integration with the IR-safe integral. This is relevant not only because it is difficult to cancel the IR divergences numerically, but also because, as for the case of $n=-3/2$ and for the $\Lambda$CDM universe, we are interested in the subleading part of the integrand, the leading part being degenerate with a counterterm of the EFTofLSS that is not calculable within the EFT anyway.

We first consider the Copter evaluation available in Ref.~\cite{Carlson:2009it}. This approach simply uses a quasi-Monte Carlo method method to separately integrate the four diagrams $P_{51}$, $P_{42}$, $P_{33}^{\rm (I)}$ and $P_{33}^{\rm (II)}$. The comparison is unfortunately straightforward, as  we were unable to get Copter to return anything but``NAN" error, which means ``not-a-number" error,
for scaling universe input, even with finite UV and IR cutoffs.  While there may be a simple fix to get numbers out for comparison, we did not find one.  

\begin{figure}[t]
\begin{center}
\includegraphics[width=0.46\textwidth]{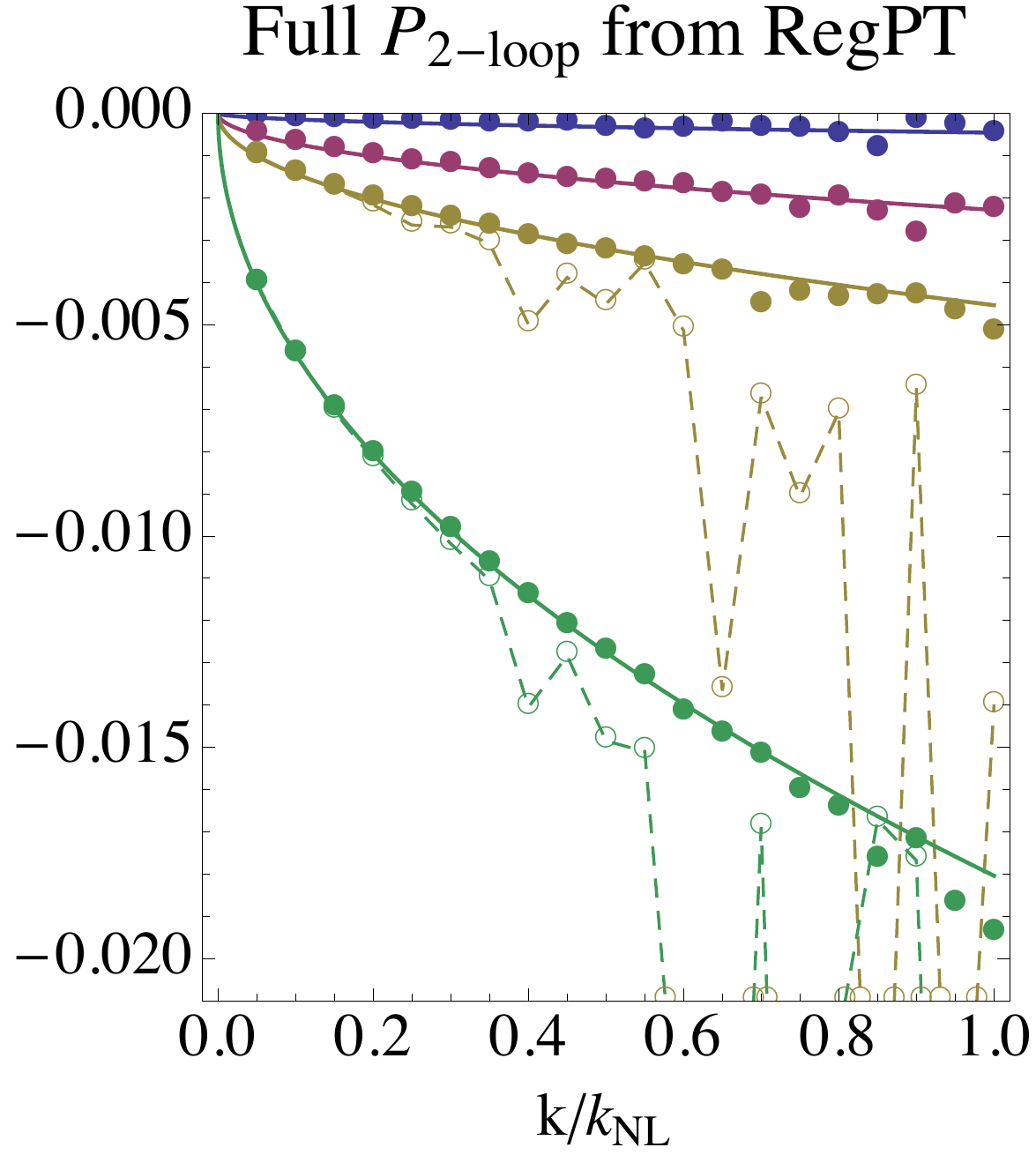}
\includegraphics[width=0.48\textwidth]{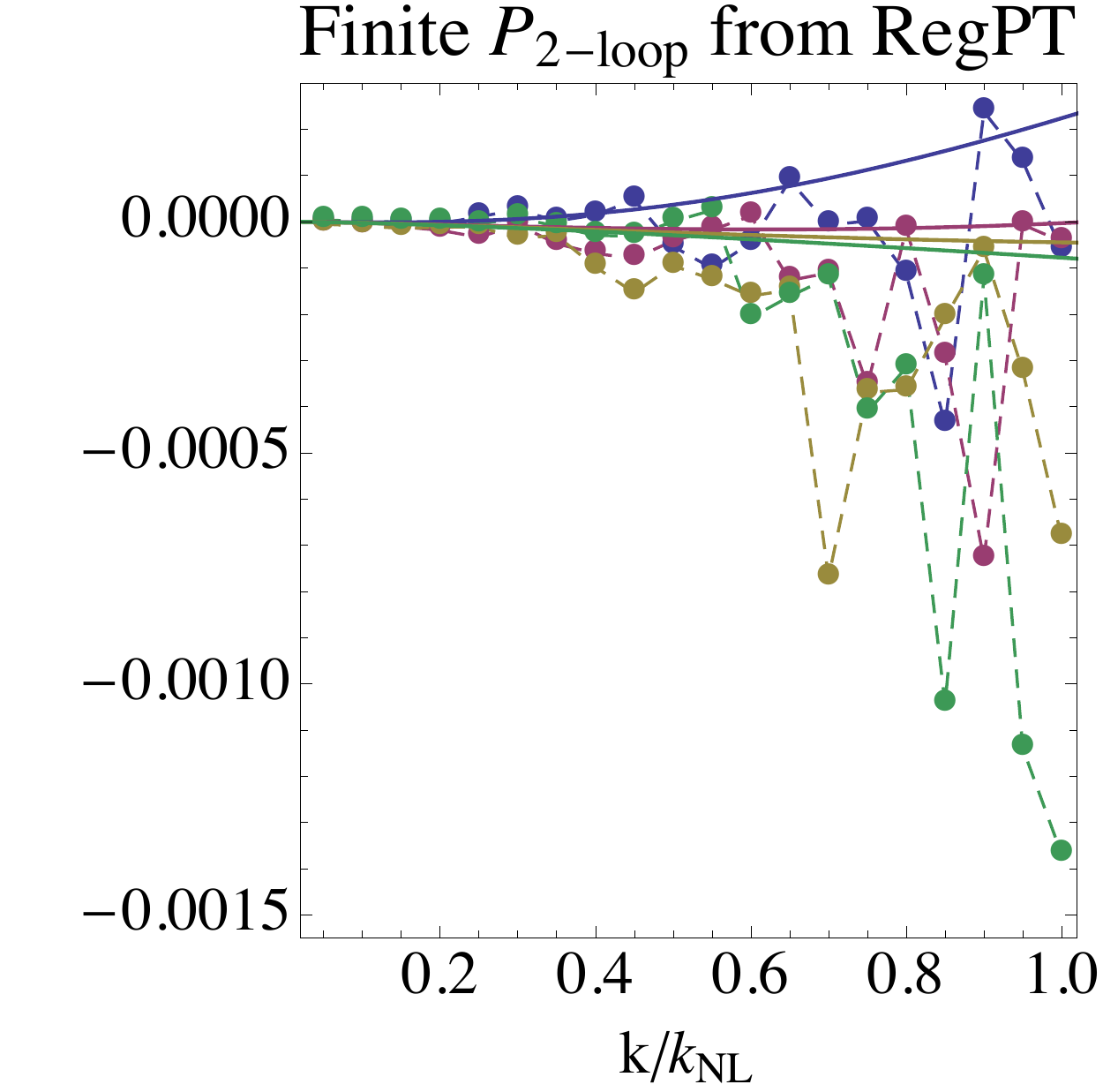}
\caption{ \label{fig:regptlow} \small\it Left: $P_\text{2-loop}$ output of RegPT for a scaling universe with $n=-3/2$, using $k_{\rm min}=0.005\knl$
 (solid points) and $k_{\rm min}=0.0005\,\knl$ (empty points).
We show $\Lambda=2\,\knl$ (blue), $10\,\knl$ (red), $20\,\knl$ (yellow), and $80\,\knl$ (green)
 and omit two small-$k_{\rm min}$ curves to avoid clutter.
The solid lines are fits to the dominant $\Lambda\, k^{1/2}$ behavior of each set of points---notice the scatter of the points about the smooth fits at higher $k$,  as well as the strong $k_{\rm min}$-dependence, signalling that IR-divergences are not properly treated.
 Right:~$P_\text{2-loop}$ after having subtracted the UV-divergent piece of each curve. The dotted lines connect the computed points as a visual aid, while the solid lines are the fitting functions from Eq.~(\ref{eq:finite}). Numerical instabilities prevent RegPT from reliably indicating the form of the terms (compare with Fig.~\ref{fig:variousLambdakmin2}).
}
\end{center}
\end{figure}

We then compare with the RegPT approach of Ref.~\cite{Taruya:2012ut}. While this is built to implement a regularised perturbation theory, one of the available outputs is the 2-loop SPT power spectrum, which is the one we are interested in for the purpose of comparison. This code does not treat the IR divergences in the way we do, but it is still able to complete calculations for scaling universes.\footnote{One should note that as of the time of writing, RegPT requires an earlier version of CUBA: (v.1.5).  We verified that the IR-safe global integrand results are unaffected by differences in CUBA versions.} 

\begin{figure}[t]
\begin{center}
\includegraphics[width=0.44\textwidth]{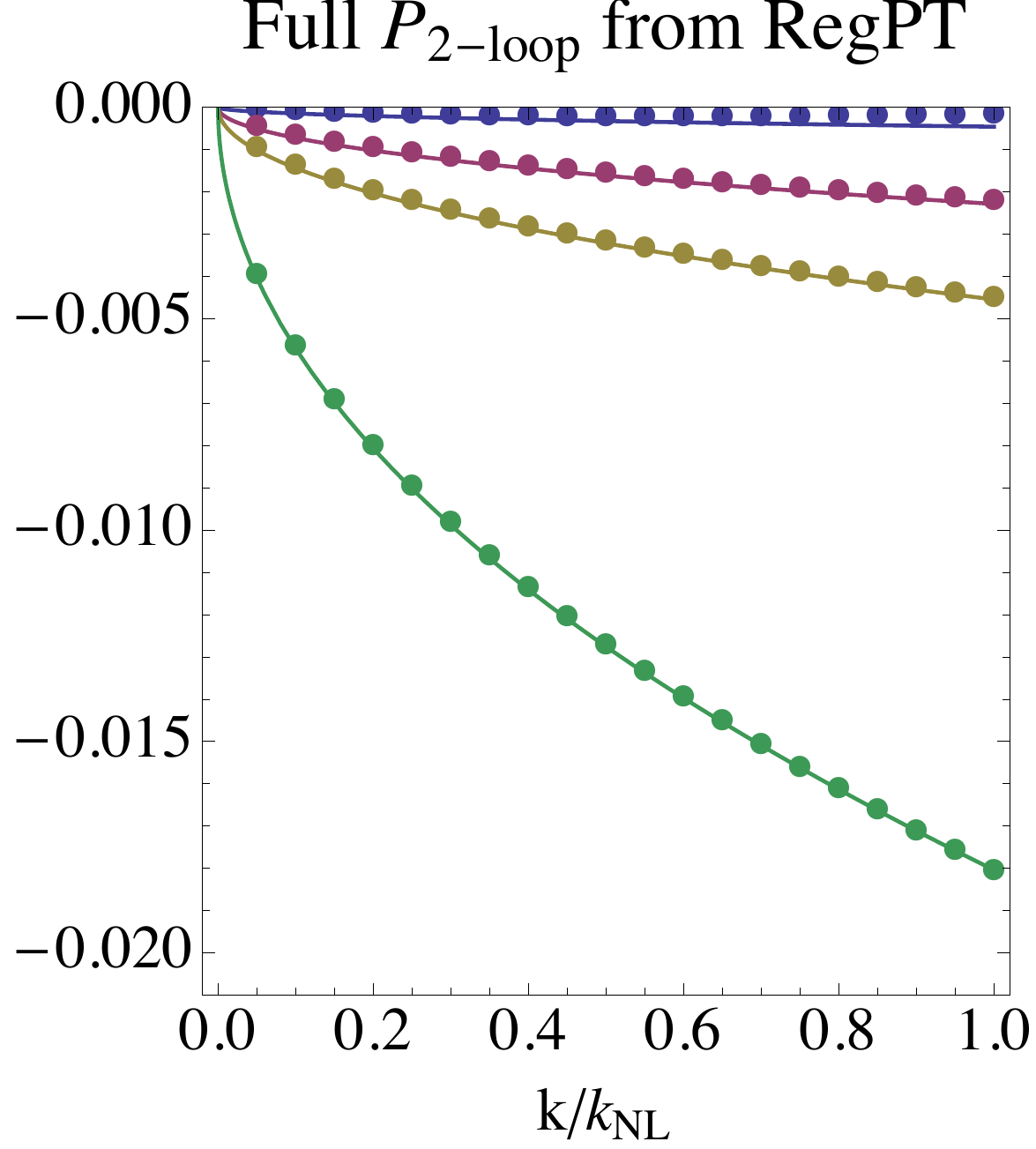}
\includegraphics[width=0.47\textwidth]{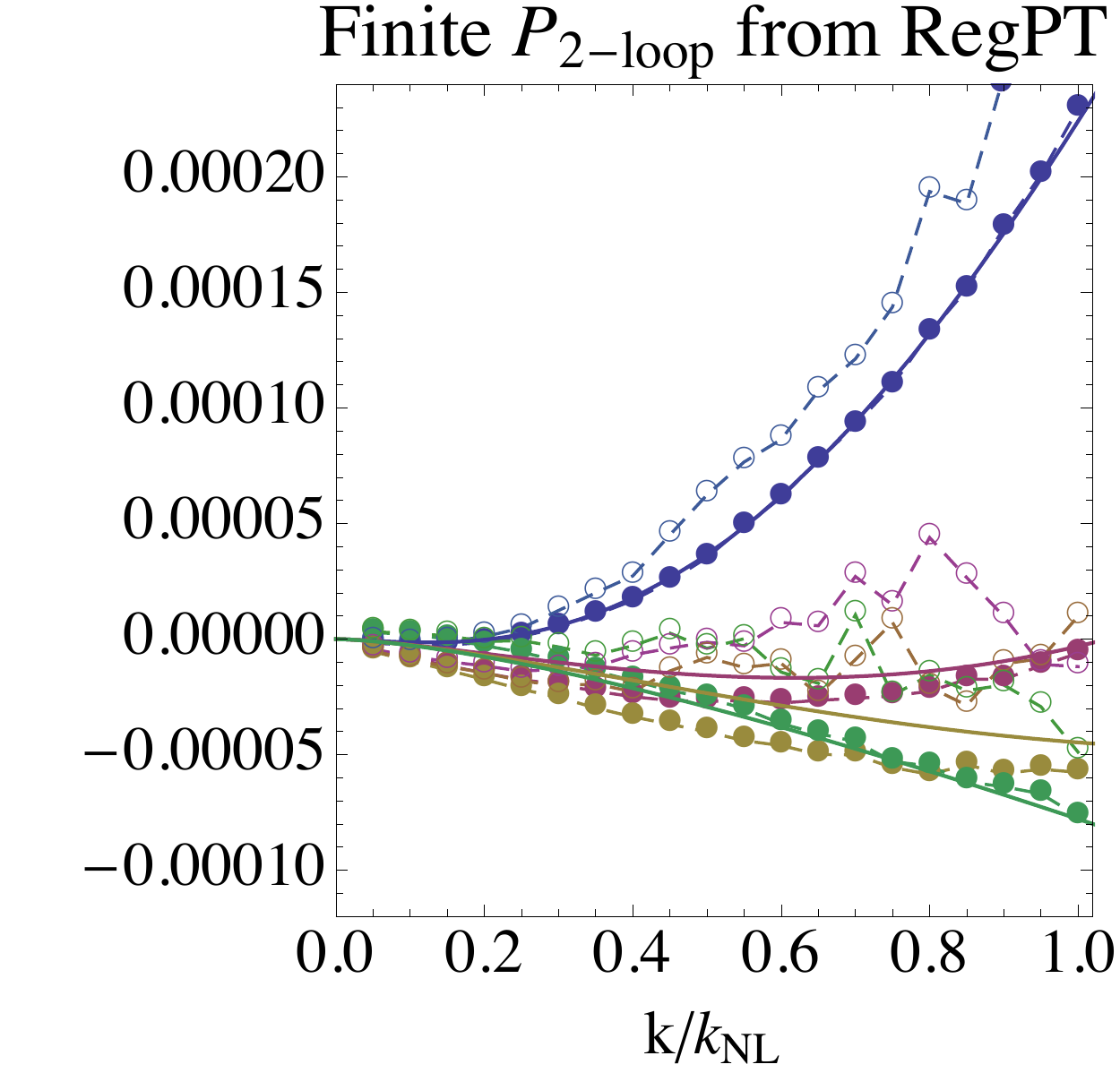}
\caption{ \label{fig:regpthigh} \small\it 
Same as Fig.~\ref{fig:regptlow}, but with increased numerical precision in the RegPT integrations, and also displaying the results for $k_{\rm min}$ reduced by a factor of 10 (empty points). There is a strong difference between the results with the two $k_{\rm min}$ values, not when we look at the full result, but after we remove the divergent part. This signals the fact that the IR-devergences are still not properly treated, and shows the importance of very high numerical precision, as we are really interested in a subleading component of the 2-loop integral. While the $k_{\rm min}=0.005\,\knl$ results are qualitatively similar to the fits to our IR-safe integrand (solid lines in right-hand frame),  there are still noticeable deviations for $\Lambda=10\,\knl$ and $20\,\knl$. Therefore, increasing the precision at the level we have done is not a robust
 solution to the IR-dependence of RegPT's 2-loop output.}
\end{center}
\end{figure}

In Fig.~\ref{fig:regptlow}, we plot the $P_\text{2-loop}$ from RegPT with $k_{\rm min}=0.005\,\knl$
 (solid points) and $k_{\rm min}=0.0005\,\knl$ (empty points) for the same values of $\Lambda$ ($2\knl$, $10\,\knl$, $20\,\knl$, and $80\,\knl$) as in Fig.~\ref{fig:variousLambdakmin2}.
With the larger IR cutoff, we observe the correct $\Lambda k^{1/2}$ behavior, but begin to see scatter of the calculated points around the smooth fit lines at higher $k$. For the smaller IR cutoff, the computation becomes unstable. Even for the larger IR cutoff, numerical noise becomes overwhelming when we cancel the UV-divergent piece with the appropriate counterterm and plot the leftover finite piece. The code's default precision is apparently insufficient to provide meaningful results for UV-convergent part of the $P_\text{2-loop}$, as evidenced by the mismatch between the numerical results from RegPT and the fitting functions obtained from our IR-safe integrand (plotted as solid lines in the right-hand frame).
 
We reiterate that the physical information of $P_\text{2-loop}$ is contained in the UV-convergent part. Therefore any numerical calculation of $P_\text{2-loop}$ for scaling universes should not only reproduce the $\Lambda$-dependence seen in the left-hand frame of Fig.~\ref{fig:regptlow} (or the corresponding version for different tilts), but should also be able to resolve the behavior of the UV-convergent piece with the wished precision.

It is also possible to increase the precision of the integration methods used by RegPT and examine the results. We show such a computation in Fig.~\ref{fig:regpthigh}, where the code's runtime has increased by a factor of $\sim$100 as a result of the heightened precision. While the results with larger $k_{\rm min}$ are qualitatively consistent with those of our IR-safe integrand (although there are noticeable deviations for $\Lambda=10\,\knl$ and $20\,\knl$), there is still strong IR-dependence that creates numerical issues even at higher than default precision. These issues become even more severe as $k_{\rm min}$ is taken lower and lower. In principle, running RegPT using arbitrarily high numerical precision would likely allow the user to recover the detailed forms of the finite part of $P_\text{2-loop}$. However, a more robust approach would implement the calculations in a way that makes IR divergences manifestly absent before the integrals are evaluated, as with our IR-safe integrand.

\end{document}